# Icing monitoring with thickness shear acoustic waves

Jaime del Moral,[a] Miguel González del Val,[b] Víctor Rico,[a] Juan R. Sánchez-Valencia,[a] Julio Mora,[b] Paloma García Gallego,[b] Francisco Carreño,[b] Andreas Winkler,[c] Agustín R. González-Elipe,[a] Ana Borrás*[a] and Stefan Jacob*[d,c,e]

[a]Nanotechnology on Surfaces and Plasma Lab, Materials Science Institute of Seville, CSIC – US, Avd. Americo Vespucio 49, 41092, Seville (Spain).

[b]National Institute for Aerospace Technology (INTA), 28850 Torrejón de Ardoz (Spain).

[c]IFW Dresden, SAWLab Saxony, 01069 Dresden (Germany).

[d]Acoustics and Mechanics, German National Metrology Institute (PTB), Bundesallee 100, 38116 Braunschweig (Germany).

[e]TU Braunschweig, Institute for Acoustics and Dynamics, Langer Kamp 19a, 38106 Braunschweig (Germany)

anaisabel.borras@icmse.csic.es; stefan.jacob@ptb.de

Icing monitoring and ice detection are critical challenges across many industries where ice formation can severely degrade performance or compromise safety. Conventional ice-sensing approaches, such as those based on surface acoustic waves (SAWs), often suffer from limited selectivity, fast saturation, and slow or energy-consuming recovery. In this work, we propose the use of thickness-shear-mode bulk acoustic waves (TSM-AWs) as a selective and robust alternative for monitoring icing phenomena, ranging from frozen sessile droplets to ice accretion in near-real scenarios within aeronautical envelopes. The sensing platform consists of a $LiNbO_3$ plate activated through lateral-field-excitation electrodes and driven electronically by a Vector or Scalar Network Analyzer. This system reliably detects both freezing conditions and temperature variations. Tracking the evolution of the magnitude of the reflection coefficient, specifically, the resonance peak and shape and frequency of $|S_{11}|$ minimum, of a strongly shear-dominant mode provides a reliable means to characterize and extract both qualitative and quantitative information about icing processes. Finite-element simulations further elucidate the physical mechanisms governing the ice monitoring capabilities. These results position TSM AW devices as promising candidates for a new generation of simple, light-weight, highly sensitive, and robust acoustic wave icing sensors.

## 1. Introduction

Icing processes are of paramount importance for a large variety of industrial activities, including aviation, ground and marine transport, electricity generation with power plants and photovoltaic panels, household elements, and refrigeration, among others [1–4]. In many of these scenarios, monitoring the environmental conditions and detecting icing events are critical to prevent fatal hazards (e.g., in airplanes or ground transport vehicles) and to ensure the smooth operation of industrial facilities, weather forecast stations or other safety services [3–6]. These application areas require the availability of robust and reliable sensor systems capable of monitoring the formation of ice and, if possible, determining the extent and characteristics of the icing events [7–10]. Moreover, to ensure continuous operation and reproducibility after successive icing events, sensor systems in these scenarios must incorporate automatic regeneration capacities to remove the accumulated layers of ice that, once reached the sensor saturation limit, render the devices useless [11–13]. These features, together with an easy implementation, simple operation, and robustness to withstand adverse climatic conditions, are requirements for the future generation of icing sensor systems. Many different approaches and procedures based on different physical principles of interrogation have been proposed for this purpose, including optical, light scattering, conductimetry and other electrical methods, mechanical cantilevers, temperature monitoring, or ultrasonic transduction [9,11,14–18].

The reliability and relatively easy implementation of AW systems for a large variety of sensing applications have been factors thrusting the investigation of this type of devices also for icing monitoring [10,19–21,21–27]. A first approach has relied on the use of surface acoustic waves (SAW), either Rayleigh, Love, or shear-horizontal SAW, generally operated in a high frequency MHz or even GHz range [20–23,26,27]. For other sensing applications, these procedures are known for their extraordinary sensitivity and, when conveniently functionalized, also for selectivity toward specific reagents [23,26,28,29]. However, their use for icing monitoring faces unsolved challenges such as response saturation, poor capacity to differentiate water from ice and, in most cases, the lack of an efficient regeneration procedure once saturation has been reached. Part of these problems appears because SAW ice sensors typically rely on the transmission characteristics of the surface wave and saturate quickly once the surface becomes blocked [26,27]. Furthermore, they require transmission paths formed by delicate interdigital transducers (IDTs), which must be located on the exposed sensor surface and are therefore subjected to harsh environmental conditions [21]. Moreover, due to the capacity of interaction of out-of-plane SAW (e.g., Rayleigh waves) with both liquid water and ice, detection of water-ice transitions is not straightforward unless less common shear-polarized SAWs are used [28,29].

Some of these challenges can be partially overcome using IDTs with a large pitch among fingers to generate Lamb-type plate waves [24,25,30,31] and other related systems for AW activation. For example, in a recent work [32], Fu et al. claimed the use of a device formed by a ZnO piezoelectric film deposited on an aluminum substrate and a pair of opposed IDTs with a relatively large pitch for activation (i.e., generating a relatively low frequency AW). Another approach recently applied by our groups has consisted of using the same set of two IDTs with a delay line configuration deposited on a $LiNbO_3$ (LN) plate, both to monitor the formation of ice generating thickness shear mode acoustic waves [31] or de-icing through the generation of Rayleigh SAWs. This dual mode activation was possible by applying specific poling strategies of the two finger combs in each IDT.

Herein, we go a step forward to selectively monitor ice accumulation and water/ice phase transformations: we propose the use of standing bulk acoustic waves (BAW) with a thickness shear mode (TSM) character that are induced by activation with a simple pair of extended electrodes in a lateral field excited (LFE) configuration [19,33–36]. Other bulk acoustic wave systems, such as high-frequency bulk acoustic resonators, are widely used in various sensing applications due to their high sensitivity [37]. Quartz crystal monitors (QCM) are another type of device, which, based on BAWs, are widely used to follow the growth of thin films [38], to detect different processes in liquid media [39,40], to monitor dew adsorption on functionalized surfaces of the device [41], or even to monitor water-ice transformations in sessile droplets [42].

The choice of TSM devices with an LFE configuration and the specific layout of the piezoelectric transducers used in the present work has been supported by several advantages. Firstly, due to the inherently high selectivity of this type of AWs, which can distinguish solid ice from liquid water based on their markedly different viscoelastic properties and on the very different penetration capacity of shear waves in each phase [34]. Secondly, by the relatively high flexibility and simplicity in the manufacturing and robust operation of transducers, triggering TSM waves, where the required LFEs are flat and consist of large patches gently pressed or glued to a piezoelectric plate, and can better withstand cycling, corrosion, scratches, or other mechanical impact events than IDTs. Herein, we have selected single crystalline 128° Y-cut plates of $LiNbO_3$, a commercially available material characterized by a high electromechanical coupling factor [43], as basic piezoelectric element for the transducers, while their activation is induced via polarization of a pair of flat extended electrodes fabricated from Cu or Al [33,35,44]. A clear advantage of the adopted design is that electrodes can be placed side down the plate surface exposed to the icing and therefore protected from the external environment. This design was used by us in a previous work dealing with AW de-icing of ice aggregates accumulated on the surface of LN plates, bare or coated with a thin film to make them hydrophobic [19]. In that investigation, we established a preliminary correlation between the icing/de-icing phenomena and the evolution of the scattering parameter $S_{11}$ vs. frequency function of the device as measured with a Vector Network Analyzer (VNA). $S_{11}$ has also been the electroacoustic parameter used in this article to directly monitor icing and elucidate the underlying physics of the operation of the LN transducers, namely, determining the characteristics of the generated vibrational wave modes and the effectiveness of AW transmission to either water or ice. The choice of $S_{11}$ as a monitoring function has been supported by a series of factors. Primarily, because in TSM resonators, it provides a sensitive and easily accessible measure of surface conditions. In addition, its natural frequency depends mostly on material properties and wafer thickness, allowing a wide range of sensor sizes without affecting the underlying mechanisms, as described in this paper. Thus, for conventional LN wafers, the frequency lies in the low MHz range, enabling inexpensive, single-channel network analyzers and avoiding interference with critical aviation bands. Moreover, the resonance is sharp, facilitating the detection and tracking of small changes with minimal hysteresis. Another advantage is that, in addition to the ice surface loading affecting both frequency and resonant peak shape (amplitude and width), frequency shifts also permit monitoring of temperature variations.

The paper is organized as follows. Firstly, to assign vibrational modes, we used a finite element model (FEM) in COMSOL Multiphysics to simulate the acoustic modes generated in the transducer over the frequency range chosen for the experiments. The identification of resonant modes with a pronounced shear character has been key in ensuring that the transducers operate on TSM waves. A similar model incorporating the effect of a water droplet or an ice aggregate on the resonant behavior of the piezoelectric plate has also been used to ensure that the transducer/selected wave mode will present a selective response to ice. Relying on these simulations, we proceed to a series of freezing-thaw cycling experiments in static conditions to verify the capacity of the device to reversibly detect phase changes. The use of AW activation to remove ice agglomerates and, thus, reuse the device for sensing is also verified at this stage. Then, we analyze the accretion of ice under dynamic conditions in an icing wind tunnel (IWT) facility replicating an atmospheric icing scenario [45]. This experiment aims at monitoring the progressive accumulation of ice over time and the recovery of the device sensing functions after saturation. Results have confirmed the suitability of TSM-AW for the selective detection of icing processes and, in the case of ice accretion in the IWT, for quantitatively monitoring the amount of accumulated ice. All these results demonstrate that LN plates excited with extended electrodes in an LFE configuration are a straightforward option for developing highly sensitive icing sensor devices with deicing capability.

## 2. Results and discussion

### 2.1 Transducer chips layout and modal analysis

The scattering parameter $S_{11}$, related with the return loss (RL=-$|S_{11}|$(dB)) is defined as

$$|S_{11}| = 20log_{10}\frac{V^-}{V^+} \qquad (1)$$

were $V^+$ and $V^-$ and $|S_{11}|$ are the incident and reflected traveling-wave voltages measured by the vector network analyzer (VNA), and the module of the complex $S_{11}$ parameter, respectively. $S_{11}$ is recorded with the experimental arrangement reported in **Figure S1** in the Supporting Information Section and exploited here as the monitoring signal to detect icing events on the surface of the piezoelectric devices. The employed device has a single port defined by two extended electrodes, and therefore $S_{11}$ is its only scatter parameter. Since it can be affected by a series of geometric variables of the system, such as small variations of LN plate thickness or the misalignment of electrodes with respect to their nominal 128° Y-cut direction, we have conducted a series of FE simulations using COMSOL Multiphysics to assess their potential impact on the $S_{11}$ resonance peaks and the reproducibility window in the responses of different chips. The details for the FE simulation methodology can be found in the Methods section and in the Supporting Information **Figures S2** and **S3.**

In these simulations, the LN system is employed as a piezoelectric model reference, considering dry, pristine LN surfaces with a crystal orientation following the 128º y-cut, with (mainly) x-propagation direction, and a nominal thickness of around 500 µm. The first evaluated parameters were the effect of possible misalignments between the crystal face cut, electrode orientation and the theoretical Euler angles of the plate surface, and the impact of minor variations in the thickness of the LN plate relative to its nominal value as supplied by the manufacturer. Simulations on the misalignment effects confirmed that variations in electrode orientation with respect to the nominal x-direction of the single-crystalline LN plate do not significantly affect the $S_{11}$ frequencies, intensities, or pattern distribution, at least for angular ranges smaller than 2º (see supporting information **Figure S4**). Indeed, the $S_{11}$ resonance peaks were more sensitive to small deviations in the orientation of the y axis from the nominal angle of 128° (Figure S4 c). These potential sources of $S_{11}$ resonance displacements, particularly the possible misalignment of electrodes, have been carefully considered during preparation of the chips. Nevertheless, the simulations showed that variations of $\pm$ 10 µm in the plate thickness around the nominal value of 500 µm may induce considerable changes in the experimental frequencies of the $S_{11}$ resonance peaks and lead to an offset of the resonant modes (see Figure S4 e), potentially affecting the experiments. Since this variable cannot be properly controlled in the laboratory, additional simulations have been carried out to fine-tune the incidence of thickness in the resonant modes of the LN plate (see supporting information, **Figure S5** and **Figure 1**). This analysis also pursues the identification of the vibrational modes present within the frequency window selected for the icing-monitoring experiments.

Figure 1 a) shows a scheme of the chip layout and an example of a FEM simulated map of displacement amplitudes (module of the displacement field) for one of the electroacoustic vibrational modes identified in the frequency range used for the detection experiments. Figure 1 b) shows a series of simulated $S_{11}$ spectra for plates with different thicknesses from 505 to 511 µm. It is apparent that the number of $S_{11}$ resonance peaks changed from two for thicknesses smaller than 507.7 µm (spectra represented in black color) to three for thicker plates (orange and green spectra). Moreover, there was a continuous decrease in the frequency of the $S_{11}$ resonance peaks and changes in their relative intensities. A more complete assessment of frequency variations with thickness is depicted in Figure 1 c). Henceforth, we will use the labels $B_n$ for these "resonance branches", with n = 1 for the lowest resonance frequency in the selected range. It is noteworthy that these branches only refer to the set of frequency/plate thickness compatible with acoustic resonances. Still, there is not a defined link between a certain branch and a specific acoustic mode (named here as $M_i$), meaning that acoustic mode characteristics can vary along the same branch. Branch $B_1$ only appears for thicknesses higher than 507.7 µm and gives rise to a resonant coupling zone for thicknesses between 508-509 µm (pale blue marked zone in Figure 1 c) where $B_1$ and $B_2$ branches almost overlap (i.e., they define a "transition zone"). It is also apparent that, after this transition zone, $B_2$ branch presents weaker resonances for higher thicknesses (fading marked with a dotted line). A deeper understanding of the evolution of lattice vibrations associated with branches $B_1$-$B_2$ in the transition and outer zones was gained by simulating the displacement amplitude maps of the vibrational modes. Figure 1 d) shows a representation of selected displacement maps for the vibrational modes associated with these two branches over the thickness range from 508.0 to 509.0 µm (see a representation with more resolution between 507.5 to 509.1 in Figure S5).

The simple observation of these maps reveals that the $B_2$ branch before the transition zone is characterized by lattice vibrations with a majority TSM character (hereafter designated as $M_1$ mode). Then, after the transition zone, the resulting acoustic mode is characterized by a predominant antisymmetric bending character typical of an $A_0$ Lamb wave [46] (designated as $M_2$ mode). We tentatively attribute the generation of this $A_0$ Lamb wave to the stress relaxation of the TSM at the two lateral borders of the plate. Interestingly, an opposite thickness dependence is found for $B_1$ branch, which depicts a $M_2$ mode before and a $M_1$ mode after the transition zone, respectively. Displacement maps in Figures 1 d) and S5 prove a full hybrid character of the wave profile in the transition zone (i.e., a $M_h$ mode), with an effective coupling and equivalent contributions of TSM and $A_0$ Lamb bending waves due to their similar phase velocity for this range of plate thickness. [47] Outside this zone, an uneven superposition of the TSM and the $A_0$ Lamb waves gives rise to modes $M_1$ (TSM predominance) and $M_2$ ($A_0$ Lamb predominance). For simplicity, when discussing in the text the sensing capacity of the TSM waves, we will refer to mode $M_1$, where this type of wave is predominant.

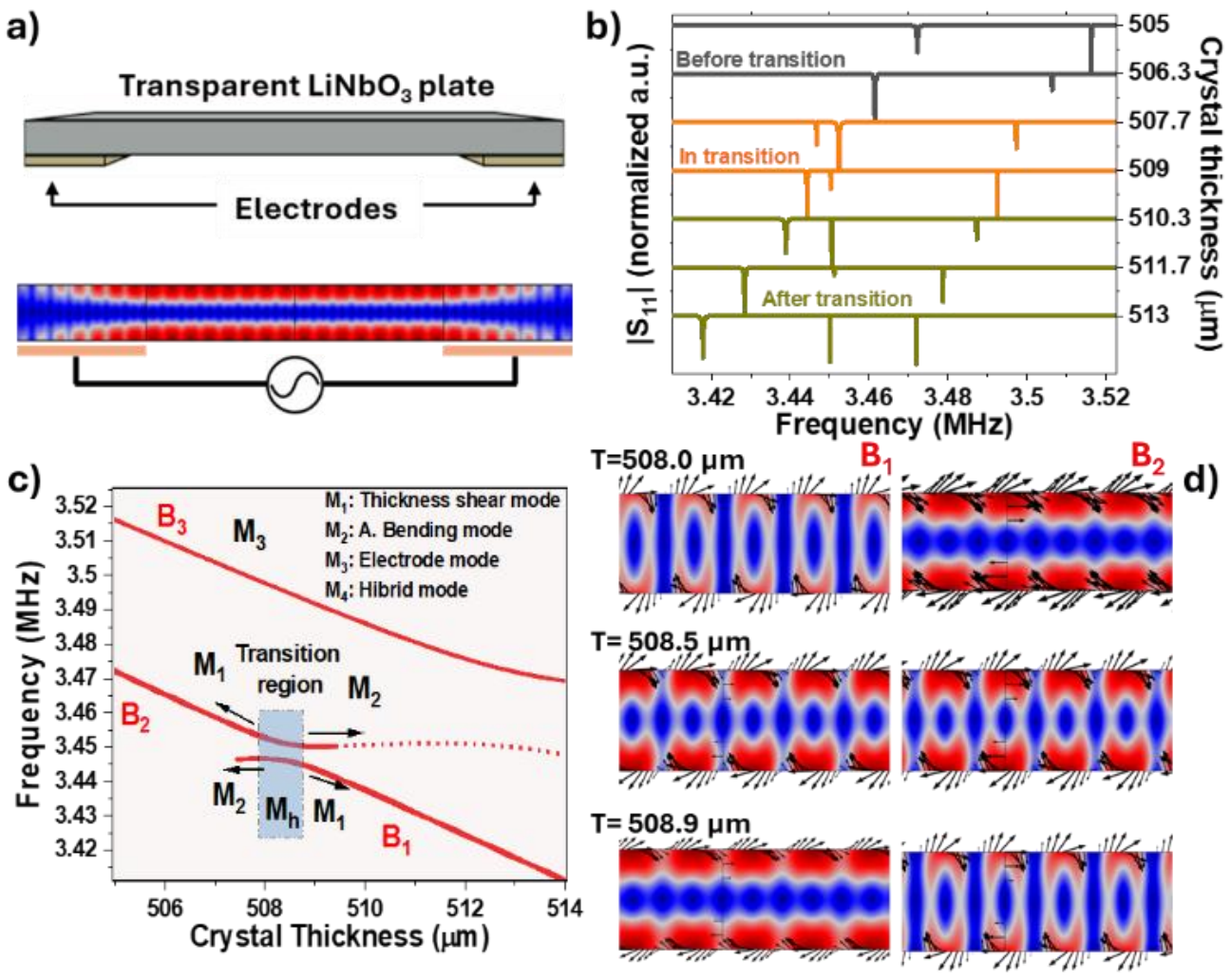

***Figure 1***. *Simulation of $S_{11}$ resonance peak distribution and vibrational properties of the plates in the chosen frequency domain as a function of thickness variations over the nominal value of 500 µm. a) Scheme of the electrode layout and a color map snapshot representing the displacement amplitudes along the whole length of the chips. b) Series of simulated $S_{11}$ spectra for increasing thicknesses of the plates. The different colors used to represent the $S_{11}$ curves reflect the three sets of $S_{11}$ resonance patterns found as a function of plate thickness relative to a transition zone. For a proper comparison of peak frequencies, the $S_{11}$ resonance peaks in each spectrum have been normalized (in the dB scale) to the most intense one taken as unity. Spectra have been vertically shifted for visualization. c) Plot of calculated $S_{11}$ peak frequencies as a function of thickness identifying a "transition zone" (pale blue zone in the plot) at around 508 µm. Branches $B_1$-$B_3$ correspond to the evolution of the three $S_{11}$ resonance peaks found in the examined frequency domain. The $B_1$ branch was absent for thicknesses smaller than 507.7 µm. Vibrational modes $M_1$-$M_2$ transform one into the other in branches $B_1$ and $B_2$ after the transition zone. The discontinuous line for $B_2$ indicates that the resonance intensity of this branch decays with the plate thickness after the transition region. d) Selected snapshots of displacement maps describing the lattice vibrations before, at, and just after the transition zone for branches $B_1$ and $B_2$. Plots illustrate the transformation of mode $M_1$ into $M_2$, and vice versa, across the transition zone. Surface displacements are represented by vector arrows.*

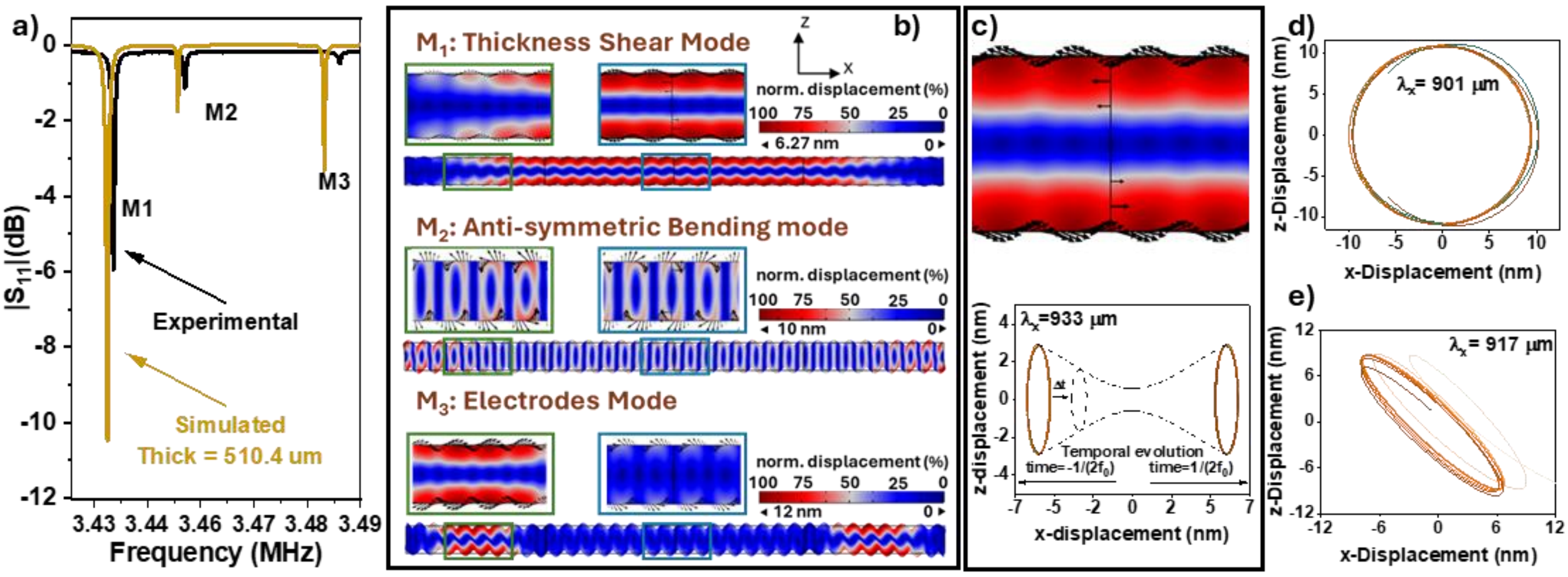

**Figure 2.** *Modal assignation analysis of the AW modes generated in the chips in the frequency range of interest. a) Comparison of experimental and in the frequency domain simulated $S_{11}$ spectra for a chip of 510.4 µm thickness and total area 1.5 x 3 cm$^2$ activated by electrodes of 1.5 x 0.8 cm$^2$. b) Snapshots of cross-section simulated displacement amplitude distributions (in colormap intensities, note the different scale in each case) for modes $M_1$, $M_2$, and $M_3$, differentiating between the regions without electrodes (central, blue-edged squares) and those where electrodes are located (edges, green-edged squares). The arrow fields in these plots represent the direction and magnitude of surface displacements for the three modes. Color scales are normalized, and the spanning range of calculated displacements is included below the scale. c) Analysis of the $M_1$ mode: enlarged view of the displacement map, and plots representing the x-z surface polarization displacements. Displacement polarization is represented by vector arrows. The predominant shear character of the $M_1$ mode is highlighted, representing in dotted lines the evolution of displacements during a vibrating period. d-e) Plots of the x-z polarization displacements of modes $M_2$ and $M_3$, correspondingly. Polarization displacement plots correspond to different times during a vibrating period and are represented on different scales. The values of $\lambda_x$ correspond to the node spatial periodicity of the surface displacement components along the x-axis*

These simulation results sustain the need to verify the distribution and intensities of $S_{11}$ resonance peaks and to proceed with an unambiguous assignment of vibrational modes for reliable sensing applications. It is noteworthy that, in the experimental validation presented below, all icing-detection experiments were conducted with crystal plates of thicknesses outside the transition zone, following the evolution of mode $M_1$. In most cases, crystals depicted a three-peak pattern (see orange and dark green color $S_{11}$ spectra in Figure 1 b) similar to that found through the simulation of LN plates thicker than 510 µm. A few experiments, though, were carried out with crystals showing only two $S_{11}$ peaks, as expected for thicknesses smaller than 507.7 µm (see the black $S_{11}$ spectra in Figure 1b).

To properly assess the experimental results, we further performed a modal analysis of the three vibrational modes observed in the $S_{11}$ spectra of a chip with a nominal thickness of $\geq$ 510 µm. **Figure 2** a) shows a comparison of experimental and simulated $S_{11}$ spectra (modes $M_1$, $M_2$ and $M_3$) for a simulated plate thickness of 510.4 µm proving a good matching both in frequency and relative intensities. Note that the simulated peaks were narrower and higher than the experimental ones (for a more detailed description of the analysis procedure, considering the Q factor see Supporting Information Section S2). In agreement with simulations, the detected modes have been identified as $M_1$ and $M_2$, and a third mode where the $A_0$ Lamb wave is predominant in the electrode-free central region and where very high shear displacements appear on the electrode lateral regions (electrode mode, $M_3$). This fair reproduction of experimental spectra demonstrates that the simulation model captures the electroacoustic activation response of the chip and accurately reproduces the lattice displacements of the three resonance modes. Snapshots of vibrational maps of the three modes are represented in Figure 2 b). An expanded view of mode $M_1$ and the resulting surface polarization displacements is shown in Figure 2 c). Displacement maps reveal that, for the three identified modes, vibration profiles differ between chip regions with and without electrodes. Focusing on the electrode-free central region, the colormap confirms that the $M_1$ mode exhibits a predominant shear character, with the most intense amplitude of displacements occurring along the chip x-direction (see polarization curve) and at the external top and bottom zones of the plate thickness (red colored bands in the map). The $M_2$ mode shows fewer differences at the electrode and free central areas of the chips, and the vibrations exhibit a bending character with intense out-of-plane surface displacements. Vibrations of the $M_3$ mode are much more intense in the electrode zones and also depict a preferential shear character, though less prominent than that of mode $M_1$. The polarization analysis of surface displacements in Figures 2 c-e) made for the electrode-free zones confirms the different shear/bending character of the three modes. Thus, x- displacements (in plane) of $M_1$ modes with a predominant TSM character proceed within a range of ± 6.27 nm for just maximum displacements of ± 2.5 nm in the z (out of plane) direction (Figure 2 c. The polarization curve of mode $M_2$ is practically a circle (Figure 2 d), confirming a considerable out-of-plane vibrational component and the bending character of this mode. Meanwhile, a preferential shear character of the $M_3$ mode is confirmed by the elliptical shape of its polarization curve (Figure 2 e). The frequencies for the three modes ranged from 3.43 MHz to 3.48 MHz with $\lambda_x$ varying from 901 µm to 933 µm, respectively. Note that these values correspond to the distance between two surface nodes mainly due to the surface contribution of the $A_0$ Lamb wave (see supporting information S6 for additional details about wavelengths of TSM and $A_0$ Lamb waves).

**2.2 Thickness-Shear-Mode AW monitoring of a liquid water to solid ice phase transition**

Thickness-shear-mode bulk acoustic waves generated in a quartz crystal microbalance (QCM) have previously been used to identify the freezing mechanism of water droplets [42]. Experiments showed the out-of-range saturation of the QCM signal upon complete icing, as well as a slightly different evolution of the resonance signal up to saturation depending on the freezing mechanism, either through bulk or contour freezing. For the purposes of the present work, these results support the use of the TSM-AWs generated in our chip device to detect transformations between water and ice. This assessment also agrees with general knowledge in the sense that, due to its liquid viscoelastic properties, especially the absence of a shear modulus, shear AWs will interact less efficiently with liquid water [29,48]. Conversely, this type of wave will interact more effectively with solid loads like ice, as long as it is in direct contact with the surface of the AW device.

To further confirm the selectivity for solid ice, we have simulated the effect of placing either a liquid water droplet or a frozen droplet on the surface of the LN plate. **Figure 3** shows the results of a FEM simulation assessing the effectiveness of mechanical energy transmission from the $M_1$ shear mode to a water droplet and to an ice aggregate with the same volume and shape. The simulation of a droplet of liquid water reveals that the $S_{11}$ resonant peak frequency and the $M_1$ mode intensity remain almost unchanged. As evidenced by Figure 3 a), the only effect is a small shift to lower frequencies, which is observable for all modes. This effect is attributable to a slight mass-damping effect due to the dynamic viscosity of water and to some heat dissipation at the interface [49]. The simulation in this figure clearly shows that the mechanical vibration of the plate is little affected by liquid water (in other words, there is poor coupling between the $M_1$ vibrations and the water and, therefore, little effect on resonance). This is further confirmed by the vibrational colormap of the system in Figure 3 b) (i.e., droplet plus plate), representing the RMS mechanical energy of the $M_1$ mode. It is apparent from this colormap that the transmission of mechanical energy from the vibrating plate to the bulk of the water droplet is rather ineffective. Conversely, a similar semiquantitative analysis carried out with ice shows that the resonance conditions change drastically. This is evidenced by the $S_{11}$ spectrum displayed in Figure 3 c), which shows a significant impact on the frequency and width of the $S_{11}$ resonance peak of the $M_1$ mode (note that the other modes are also affected). We deem that during freezing, the load impedance changes drastically as a result of the phase transition from liquid to solid, thus modifying the coupling conditions between the load and the piezoelectric plate. This, in turn, affects the resonance conditions of the system. As shown in Figure 3d), the coupled system exhibits a resonance pattern that differs significantly from that of the bare TSM, with vibrations occurring not only in the piezoelectric substrate but also within the bulk of the ice droplet. This is further assessed in Figure 3e), which compares the profile of accumulated mechanical energy along the central radius (z-axis inside the droplet) within the water and ice droplets. The inset in this figure shows that in the case of the water droplet, energy density abruptly decreases within the first 300 µm from the LN surface due to viscous dispersion [50]. This means that most part of the droplet volume does not interact with the acoustic wave. In contrast, for ice, the $M_1$ mode is able to penetrate and spread within the aggregate to be reflected by its outer boundary. As a consequence, standing acoustic waves form within the ice domain, defining the node structure of the energy-density domains reported in panels d) and e). This difference in behavior between AW transmission in water and ice strongly supports the exploitation of TSM AWs generated in the LN chip devices for the selective detection of icing events.

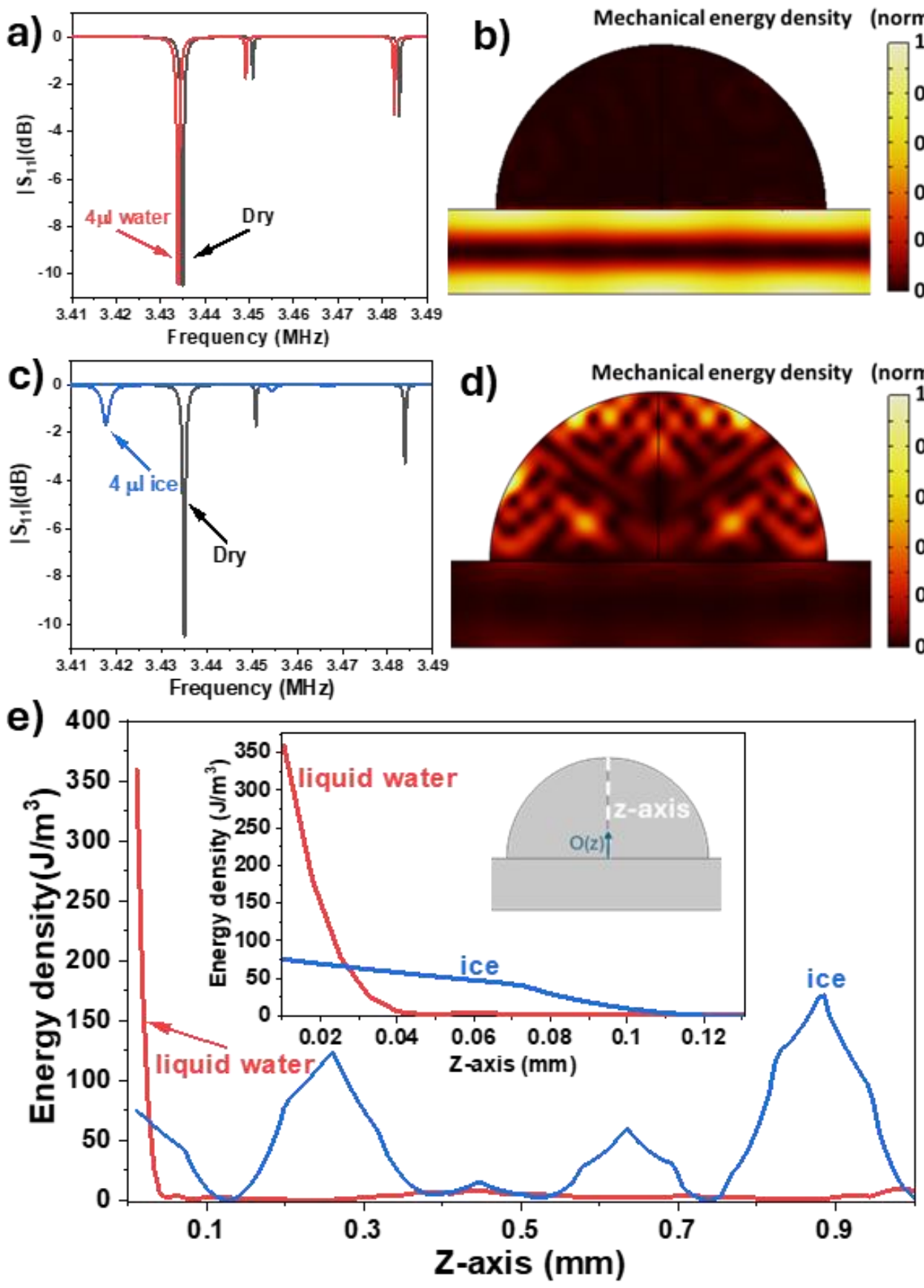

**Figure 3.** *FEM of the interaction of TSM-AW $M_1$ mode with water and ice droplets placed on the surface of the LN chips. a) Simulated $S_{11}$ spectra with and without a water droplet (4 µl) on the plate surface. b) Color map of the energy density per volume unit both in the water droplet and the LN chip. c) Simulated $S_{11}$ spectra with and without an ice aggregate with the same shape and size as the water droplet placed on the surface of the chip. d) Color map of the energy density accumulated-per volume unit, both in the ice aggregate and in the LN plate. Note that dark coloring in the color scale represents little or no accumulation of energy. e) acoustic energy density distribution along a central radius (white straight line in the droplet scheme) inside the water droplet and ice aggregate. The inset zooms out the region close to the interface. The z-axis value indicates the distance along the chosen radius between each evaluated point and the interface.*

### 2.3 Detection of water-ice phase changes in static conditions

To prove the ice monitoring capacity of the chip devices, we carried out a first experiment in static conditions following the evolution of the $M_1$ mode of the $S_{11}$ spectra when a droplet of water transforms into an ice particle and back, i.e., following a freezing-thaw phase transformation. Results reported in **Figure 4** correspond to a freezing experiment carried out with a water droplet (4 µl) placed at room temperature on the surface of the chip, which was then progressively cooled down to the set temperature of -17 ºC (see Methods for experimental details). Selected $S_{11}$ spectra in Figure 4 a) were recorded with the VNA after placing the droplet on the surface of the chip and thereafter following the cooling progression while water was still in liquid form (dark blue $S_{11}$ spectra in this figure). This series of spectra depicted practically no change in intensity and a very slight variation in frequency. Please note that the spectra recorded at different times have been vertically displaced for visualization. Time zero corresponds to the first moment when the icing process is detected by the $S_{11}$ signal. Therefore, $S_{11}$ spectra associated to negative times correspond to experimental stages with the droplet in a liquid supercooled state [51,52] as verified by direct observation (see the selected snapshots gathered in Figure 4 c); a video recording of the whole experiment can be seen in the supporting information, **Video S1**). Images show that freezing occurs through a common heterogeneous mechanism, with solidification triggered at the interface with the substrate and progressing upwards to complete solidification. Remarkably, while the droplet still consisted of a heterogeneous mixture of solid ice particles and liquid water (i.e., during the freezing process), the $S_{11}$ spectral shape varied significantly, giving rise to new resonant features as illustrated by the red line $S_{11}$ spectra in Figure 4 a). The observed changes, characterized by less intense and wider $S_{11}$ resonance peaks agree with a heterogenous freezing mechanism where the thin ice layer initially formed at the interface provides a more efficient viscous coupling mechanism between the droplet and the vibrating crystal at resonance. A general view of the $S_{11}$ evolution can be found in the peak-intensity colormap in Figure 4 b), where both frequency and intensity of $S_{11}$ resonance peaks are represented as a function of time. This plot reveals that the crystal resonance oscillates and weakens during the freezing transition when the initial well-defined peaks (dark blue spectra) begin to shatter into an apparently chaotic regime of unstable and weak resonances (red curves in Figure 4 a) and zones ii and iii in Figure 4b)). This extreme variability in the resonant peaks pattern persisted throughout the freezing process, but stabilized once the entire droplet had solidified. The images in Figure 4 c) are characteristic of selected stages of the experiment up to its final frozen state (see figure caption). When the droplet was completely solidified, the $S_{11}$ spectra showed a new pattern, as illustrated by the light blue signals in panel a) and by the zone designated as iv in the color map of panel b).

It is noteworthy that although the evolvement of the $S_{11}$ spectra during freezing was unpredictable, the distribution of resonance features after freezing followed a predictable trend from one experiment to another. Thus, a common result across various experiments (e.g., changing the cooling rate or the water droplet volume) was that the lowest frequency and most prominent $S_{11}$ resonance peak, appearing once complete icing was achieved, had a slightly higher frequency than the original $S_{11}$ resonance peak of the $M_1$ mode. From now on, we will designate this new peak as $M_1$*, without implying any specific assignation of vibrational characteristics. This $M_1$-to-$M_1$* conversion was consistently reproduced when working with small ice aggregates (1-10 µl), and we retain it as a simple and specific experimental signature of a liquid water-ice transformation in sessile water droplets. However, the magnitudes of the frequency shift and intensity drop of the $S_{11}$ features varied significantly between, in appearance, equivalent experiments. It could thus be realized that minor changes in experimental parameters such as the position of the droplet, cooling rate and humidity level or the final geometry of the ice droplet had a strong impact on the resonance pattern during transition.

To test the reversibility of the monitoring process, a complete freezing-thaw cycle experiment was carried out under the same set of conditions as for the experiment in Figures 4 a-c). Plot in Figure 4 d) shows the evolution with temperature of the relative frequency shift expressed as the ratio between the frequency parameter of the $M_1$ peak with respect to $f_0$, its frequency at 5 ºC (i.e. $\Delta f/f_0$). During cooling (dark blue dots and line) $M_1$ frequency mode drifted linearly to higher frequencies. Afterwards, the supercooled droplet remained liquid for several minutes at -17 ºC before the freezing transition began producing a sudden increase in frequency consistent with the previous experiment in Figures 4a-c (red line on the left side of the plot, labelled as "transition"). After this transition, the system was left to stabilize for thirty minutes. Then, the reverse thaw process was also continuously monitored following the evolution of the $M_1$* peak for increasing temperatures until the melting of ice. Noticeably, after a little transient regime (i.e., the observed small frequency shift to higher frequencies at -13 ºC) no noticeable changes in intensity or shape of the $S_{11}$ spectra occurred during the rest of the heating process, except for a continuous red shift in the $M_1$* frequency, which ran in parallel to the M1 peak frequency shift during cooling (c.f. see light and dark blue dot and lines in Figure 4 d). Interestingly, both lines converge after ice melting, as signaled in the plot by the red melting transition line. We attribute the similar slopes for the linear changes in frequency either during the cooling or the heating processes of the chip, to the natural variation of resonance properties of LN with temperature. This variation is an intrinsic property of $LiNbO_3$ and is described by its Temperature Coefficient of Frequency [53], a parameter that, in the current experiment, can be derived from the slope of the two linear sequences of points in Figure 4 d). The determined value of -86.11 ppm/ºC is in good agreement with reported values of this coefficient (e.g., − 86.72 ppm/°C in literature [53]). From a practical point of view, this correlation between $M_1$/$M_1$* peaks frequency and temperature supports the use of calibrated chip devices to follow temperature variations in real applications.

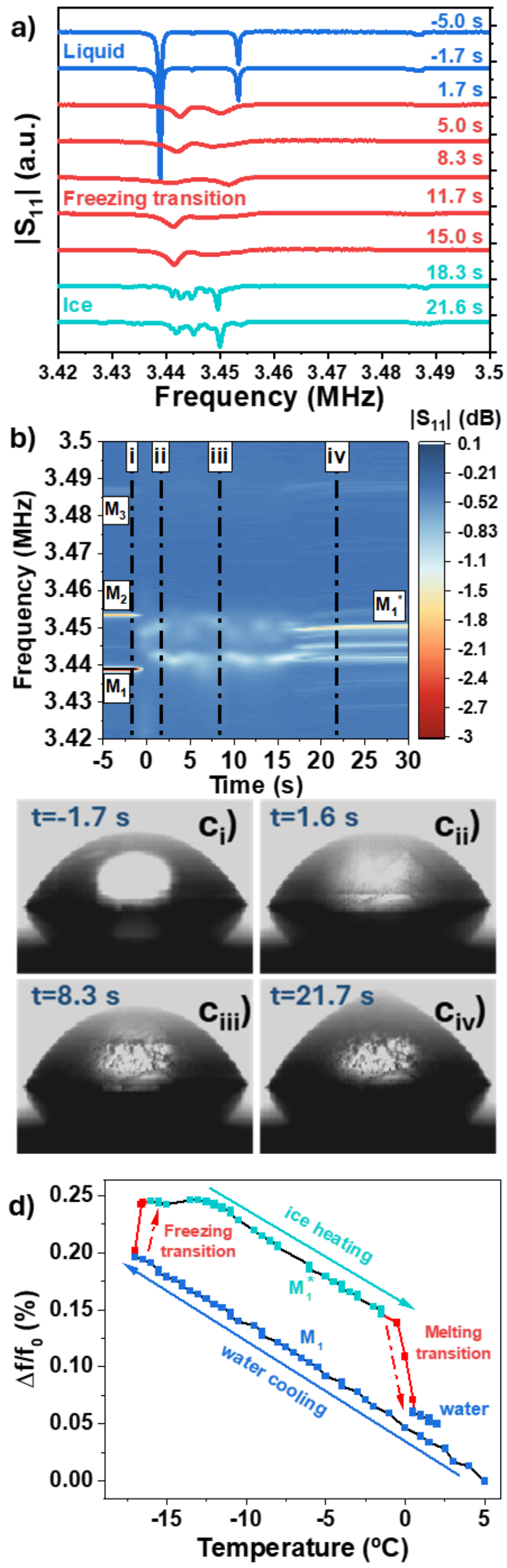

***Figure 4.*** *Monitoring of freezing and thaw processes of 4 µl water droplets. a) Selected $S_{11}$ spectra recorded at different times during cooling in the chamber for a chip with a water droplet placed on the surface of the LN plate until its complete solidification. Spectra recorded at different times have been vertically displaced for visualization. b) Colormap reporting the evolution of the intensity and frequency of $S_{11}$ peaks during the freezing process. Dashed-dotted lines highlight the cooling (i), freezing (ii, iii), and completely frozen (iv) stages for the three resonance modes $M_1$-$M_3$ as labelled. Time zero has been taken at the moment of detecting the beginning of the freezing process. c) Snapshots of the water droplet during the freezing process. The images correspond to a supercooled water droplet ($c_i$), droplets during the freezing stage ($c_{ii}$ and $c_{iii}$), and the completely frozen droplet ($c_{iv}$). Time zero corresponds to the moment when the icing process was first detected. d) Evolution with temperature of $M_1$ and $M_1$* relative frequency shifts (i.e. Δf/f0) during the cooling and heating of a chip with, respectively, a droplet of water or ice on its surface. The plot is divided into four zones referred as chip cooling, freezing transition, chip heating, and melting transition. Dots corresponding to the freezing, heating and melting transition stages represent the frequency of the most intense peak of each $S_{11}$ spectrum recorded.*

The droplet used in the previous experiments was small (i.e., 4 µl) and, once deposited on the LN substrate, covered a zone of approximately 7.83 mm$^2$, defined by a circumference of 3.15 mm diameter for a wetting contact angle of 60º. To assess the impact of water volume and droplet distribution on the chip surface, we conducted two additional experiments. The first one consisted of dripping six small droplets of water at ambient temperature (4 µl each), forming a regular pattern as indicated by the insets in the left-hand panels of **Figure 5** a). Then, we tracked the evolution of the $S_{11}$ resonance peaks as the chamber temperature decreased. In almost all repetitions of this experiment, freezing of water droplets proceeds bottom to top (i.e., heterogeneous freezing), in a sequential way, which was random and unpredictable. Figure 5 a) reports the evolution of $S_{11}$ spectra ($M_1$ and $M_3$ modes) recorded after the completion of the sequential freezing of the water droplets. It is noteworthy that the peak frequencies are shifted to higher values with respect to the curves in Figure 4 and that no mode at low frequencies associated to the $B_1$ branch appear in the spectrum as expected for a crystal with a thickness smaller than that defined by the transition zone (c.f., Figure 1 c); further simulations to reproduce the experimental $S_{11}$ curves confirmed that the employed crystal should have a thickness of 474.5 µm (see Supporting information S7). Both $S_{11}$ curves and colormaps representations as a function of time are included in the left-hand and right-hand sides of Figure 5 a), respectively. The colormaps provide a clear view of both the frequency variation of the $M_1$ peak and the appearance of new peaks and resonance features over a wide range of frequencies upon cooling and freezing. These spectra reveal little change after the freezing of the first water droplet located in this particular experiment at the upper-left corner of the droplet array. In comparison with the experiment in Figure 4 for one droplet freezing where the droplet was located at the center of the chip, this experiment proved that droplet location onto the chip surface is an important parameter controlling the $S_{11}$ changes during the icing events of small aggregates.

Upon progression of successive freezing events, $S_{11}$ spectra dramatically changed in intensity, number of peaks/new features and frequency. A comprehensive description of these changes up to the freezing of the six droplets is reported in Figure 5 b) in the form of a single colormap. The observed changes in Figures 5 a) and 5 b) reveal a drastic decrease in the intensity and a widespread distribution of $S_{11}$ features already for the third and fourth droplet freezing events (for the sake of clarity, in Figure 5 a), no color diagrams are included for the last two freezing processes, where no well-defined $S_{11}$ features could be monitored).

Significantly, a similar effect consisting of the disappearance of defined $S_{11}$ resonance features was observed in a second experiment where a large amount of water (1.5 ml) was dripped to almost cover the whole surface of the LN chip, and then frozen (see the results of this freezing experiment in **Figure S8**). A common trend in these two experiments is that even when a relatively large mass of water is distributed over a large portion of the surface of the LN chip, the $S_{11}$ spectra are little affected, and therefore freezing can be effectively correlated with the disappearance of well-defined $S_{11}$ signature(s). In other words, although a detailed mechanistic description of the freezing sequence is not accessible by monitoring the $S_{11}$ resonance peaks, the freezing transition event(s) can be unequivocally detected independently of the amount of ice formed on the surface.

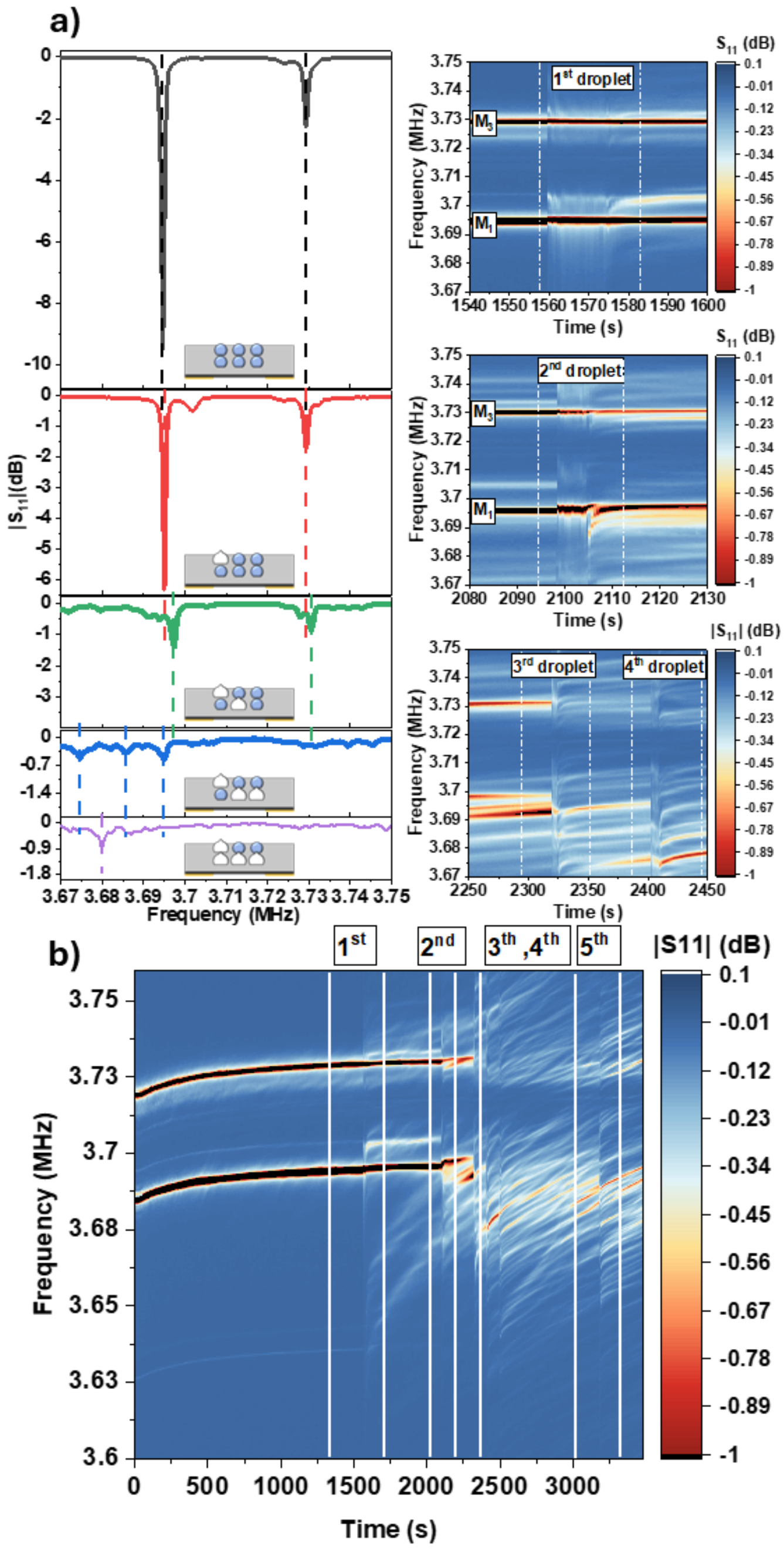

***Figure 5.*** *Monitoring the successive freezing events of water droplets on the surface of the chip. a) (left) $S_{11}$ curves recorded for an LN device chip with six liquid water droplets distributed onto its surface as indicated in the insets. From top to bottom, the panels correspond to the curves recorded after the successive freezing of 1 to 4 droplets as indicated; (right) Colormaps describing the evolution of the $S_{11}$ spectra as a function of time. The four reported colormaps depict the evolution of the $S_{11}$ spectra as the first four droplets freeze. Note that the time scales are different for each freezing event. b) Comprehensive colormap describing the evolution of $S_{11}$ spectra from the beginning of the experiment to the freezing of the sixth droplet. The white lines in the colormaps indicate the freezing of successive water droplets, as indicated.*

**2.4 Transducer chip recovering using AWs: an efficient counter saturation approach.**

From a practical perspective, an additional advantage of using TSM-AWs in LN chips for ice sensor development is that the same AWs used for detection can be applied for ice removal once the chip response saturation level is reached. This possibility would enable long-term monitoring of icing events by systematically removing the ice accumulated after specified periods. The process requires powering the piezoelectric transducer with a high-intensity activation signal at the resonant frequency of one of the remaining $S_{11}$ peaks (e.g., the M1* peak) after ice formation. Such approach has already been used by us in ref. [19] . Herein, we compared the evolution of the $S_{11}$ spectra after freezing a water droplet by either warming the device with the Peltier plate or activating the thaw process with AWs with the LN chip at minus zero temperatures. Results reported in **Figure 6** a) show the expected transformation from $M_1$ to $M_1$* after the freezing of the droplet. Then, melting the ice aggregate (c.f., Figure 6 c) by directly heating the chip device restored the initial $S_{11}$ spectra, albeit inducing a general shift to lower frequencies, attributable to an increase in chip temperature. A temperature difference of 23.8 ºC can be observed between the state of the chip before freezing, and its state after melting the ice. The chip is now prepared for a new cooling process and is ready to detect another freezing event.

It is worth stressing that direct heating of the LN chip is efficient but requires additional equipment, is time-consuming, and may compromise reproducibility (i.e., the chip surface temperature would remain higher than the ambient temperature until heat dissipates and the temperature equilibrates with the environment). Heating may also induce uncontrolled thermal gradients and accelerate material fatigue, leading to micro-cracking or delamination over repeated cycles. To circumvent these problems, we propose a reciprocal approach based on the same physical principles used for sensing. In this approach, energy is supplied directly to the system through the sensor ports and transformed into heat through friction inside the ice. Figure 6 b) shows the M1* to M1 evolution after the activation of the LN plate with a 4 W electrical inlet power to induce the melting by TSM AWs of the ice particle (see snapshots in Figure 6 d) and the **Video S2**) (note that the crystals employed for the experiments in Figures 6a) and b) came from different batches and did not have the same $S_{11}$ original spectra). Since, according to our simulation analysis, the AW energy density concentrates inside the ice aggregate (c.f., Figure 3 d), this method is efficient in transforming the ice into liquid water (and even to aerosolize the droplet, data not shown), as demonstrated in previous experiments for energy-efficient ice removal [19]. The system can therefore be used as a multifunctional sensor–actuator platform without requiring modification of the substrate or its surroundings.

It is noteworthy in this regard that comparing the $S_{11}$ spectra of the chip with the water droplet before freezing with those of the chip after AW-induced ice melting reveals almost complete reproducibility in spectral shape and intensity, demonstrating the recovery of the sensing capacity for monitoring a new freezing event. This reversible behavior also supports the claim that the transducer response is not affected by hysteresis, i.e., it does not depend on its previous use or manipulation but rather on the surface state and environmental conditions.

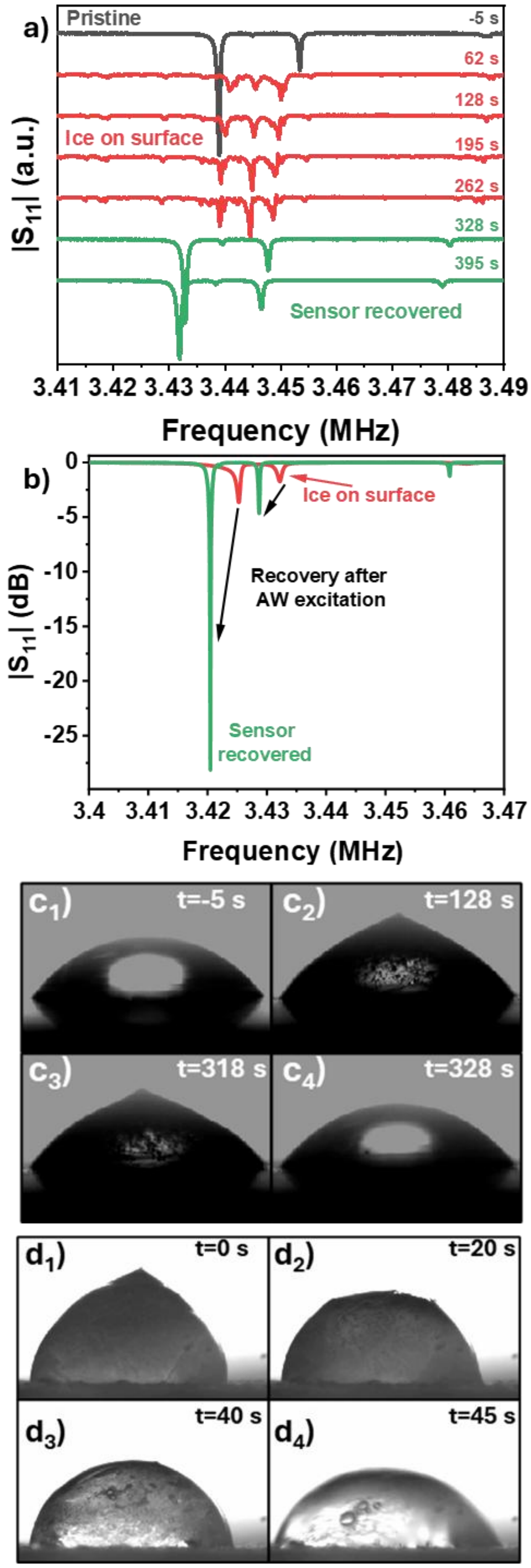

***Figure 6**. Recovery of the sensing capacity of the chip upon melting of ice aggregates. $S_{11}$ spectra recorded before and after freezing a water droplet on the surface of the chip and the recovery of the initial state upon melting by heating (a) and AW activation (b). Snapshots of the water droplet in the initial state ($c_1$ and $d_1$), as ice after freezing ($c_2$/$c_3$ and $d_2$/$d_3$), and in the form of water after melting ($c_4$ and $d_4$) by either heating ($c_4$) or AW activation ($d_4$). Time zero corresponds to the moment when the icing process was first detected by the $S_{11}$ signal.*

**2.5 TSM-AWs to monitor ice accretion under dynamic conditions.**

The previous results obtained at laboratory scale have shown the capacity of the chip devices to track water–ice transitions in droplets and to recover the transducer response upon AW activation. In this section, we investigate the behaviour of the system under more realistic conditions, namely upon impingement of supercooled small droplets under high-wind velocity as required for aeronautic applications. Under these conditions, ice gradually accumulates on the chip surface as microdroplets freeze upon impact, progressively increasing the ice layer thickness. This causes the selected parts of the chip to rapidly become ice-covered, thereby stress-testing the resolution and responsiveness of the sensor system.

Ice accretion was monitored in experiments carried out in an Icing Wind Tunnel (IWT) at wind velocities up to 50 m/s and temperatures around -15 ºC (see additional experimental conditions in the Methods section). A critical issue in this experiment was to define a precise ice accretion area on the surface of the chip and to ensure that accretion proceeded homogeneously over this area. Such a control over the accretion process was provided by a collimator placed in front of the chip surface. Details of this arrangement are provided in ref. [19]. The results of a characteristic ice accretion experiment under such conditions are reported in **Figure 7**. The pattern evolution of the $S_{11}$ spectra (highlighted in panel a) in red) was similar for a large number of tests carried out under similar conditions. It is characterized by a progressive decrease in the $M_1$ frequency and a progressive broadening and decrease of peak intensity. The appearance of some additional spurious low-intensity $S_{11}$ peaks may occasionally accompany the observed changes in the $M_1$ peak. In no case was this a complex pattern, such as those formed when monitoring either a single or multiple droplet freezing processes (cf. Figures 4 and 5). Differences in the icing process mechanism can account for these differences, as explained in section 2.6. The colormap representation in Figure 7 b) confirms that the $M_1$ peak continuously shifts and decreases in both frequency and intensity with ice accretion time. This behavior can be taken as a fingerprint of the homogeneous accretion of a progressively thicker layer of ice cast by the collimator window. The well area controlled of the accreted ice layer is demonstrated by the cross-section images reported in Figure 7 c) and **Figure S9,** which were taken in an "ad hoc" experiment exposing the chip for progressively longer times to IWT air flows with a very high liquid water content (LWC) of 0.4 g $m^{-3}$. As detailed in the Methods section, it is possible to estimate the accreted ice mass per unit surface area, defined in this work as the Mass Surface Density (MSD). In particular, the MSD estimated at the end of the experiment in Figure 7 a) was 0.05 mg·$mm^{-2}$, causing a relative $M_1$ frequency displacement of -6 %. This MSD value corresponds to an equivalent thickness of 0.04 mm, assuming a density of approximately 0.8 g·$cm^{-3}$ for the mixed ice, which is expected to form under the conditions of the experiment [54].

The representation of $\Delta f/f_0$ for the $M_1$ peak as a function of MSD(t) defines a curve that can be adjusted with an exponential function (c.f., Figure 7 d). This renders a lineal representation of both the experimental and fitted $\Delta f/f_0$ vs MSD curves when plotted on a logarithmic scale, which can be useful for practical purposes (see supporting information **Figure S10)**. However, it should be stressed the pure instrumental character of this representation and that it does not imply any specific physical interaction mechanism between the AWs and the ice. In fact, a proper fitting was also possible using a polynomial function as commonly used to describe the frequency evolution of QCM with the accumulated mass of films deposited onto its surface. [55]

A noticeable difference of this experiment with respect to those in previous sections of this work is the assumption of a "pre-stress" effect on the $M_1$ frequency, which would be introduced by the holder employed in the IWT experiments (see Methods section 4.3.3 and supporting information SI 1). The need to fix the plate to prevent its fall during the experiment introduced this initial frequency shift from the natural free-standing frequency of the crystal, $f_{free}$, to the experimental initial frequency of the mounted chip, $f_0$, which results to be smaller than the former (this displacement is marked in blue in Figure 7d). In practical terms this shift is equivalent to that of a virtual mass loaded onto the crystal, whose effect is circumvented by just taking the "pre-stressed" $M_1$ frequency as reference.

To better assess the chip behavior and the interaction between the $M_1$ TSM-AW mode and an ice layer on the chip surface, we carried out a FEM simulation that mimics the formation of a homogeneous ice layer on the central part of the chip. Figure 7 e) shows a snapshot of the calculated $M_1$ displacements map for the chip covered by a thin (approximately 0.1 mm, equivalent to an MSD of 0.08 mg mm$^{-2}$) and homogeneous layer of ice with the same lateral dimensions as in the experiment. Interestingly, the colormap shows that the lattice vibrations of mode M1 are efficiently transmitted to the ice layer (note the large amplitude of displacements affecting the ice layer and that the red/black colorations in the ice layer follow the profile found for the lattice displacements in the chip plate). This simulation result supports the conclusion that the chip-plus-ice-layer system behaves as a single oscillating, well-coupled device.

The evolution of the $M_1$ peak frequency was also calculated by FEM analysis for increasing MSD values of accreted ice (i.e., for a progressively thicker layer of ice on a chip surface area as defined experimentally by the collimator). Calculations were carried out for a free-standing chip, whose original frequency has been called $f_{free}$. The curve obtained in that case presents a decaying shape consisting of two well differentiated zones: a first linear decaying zone and a sharp decaying zone that could be fitted with an exponential (see supporting information **Figure S11**). Interestingly, as shown in Figure 7f) a reasonably good fitting between the experimental $\Delta f/f_0$-MSD and simulated $\Delta f/f_{free}$-$MSD_{free}$ curves was possible when disregarding the initial lineal variation found in the latter. This zero adjustment for the simulated curve can be taken as equivalent to the experimental effect produced by the pre-stress introduced when fixing the chip. This suggests that experiment and simulation are reproducing equivalent processes and that the chip transduction mechanism can be well described by the AW interactions displayed in Figure 7e). Therefore, we tentatively suggest that the missing linear decaying zone in the experimental curve might be related to the initial decrease in frequency found when mounting the chip in the holder. In other terms, the small pressure exerted by the four spring pins used to keep the plate fixed produces a decrease in frequency that is equivalent to an initial accumulation of mass in the simulation.

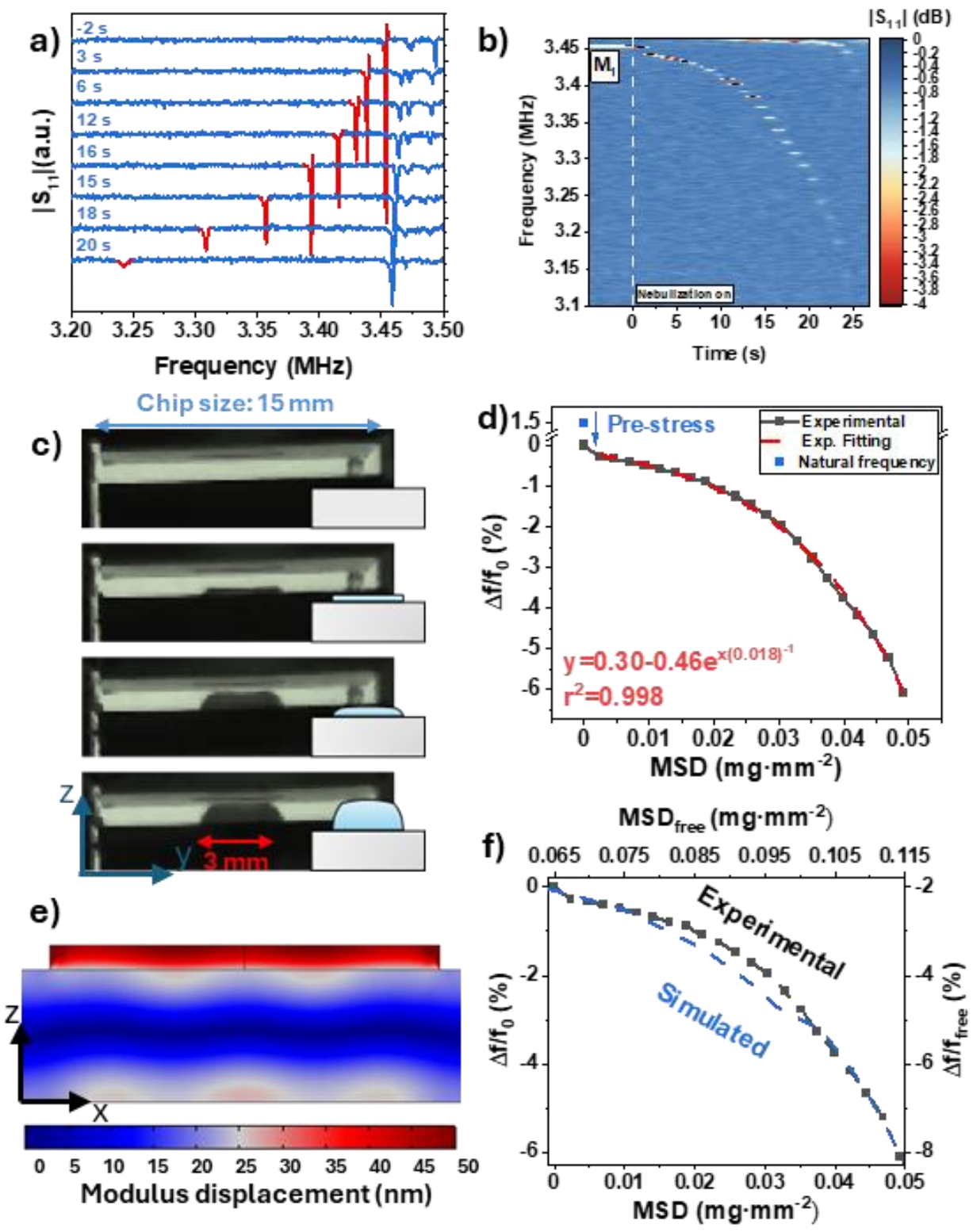

**Figure 7.** *Monitoring the ice accretion under dynamic conditions with TSM-AWs. a) Evolution of $S_{11}$ spectra with ice accretion time. The peak corresponding to the $M_1$ mode, which is used for monitoring purposes, is highlighted in red. b) Colormap reporting the evolution of the $M_1$ mode with the accretion time. c) Snapshot images along the y-axis taken (from top to bottom) after progressively longer exposure times of the device to air flows with 0.4 g·m$^{-3}$ LWC. For clarity, schematic drawings of the lateral views of the accumulated ice on the area of the chip cast by the projection of the collimator aperture.* d) *Relative variation of the frequency of the $M_1$ mode vs. the MSD and exponential function used to reproduce the experimental curve. The dots corresponding to the IWT experiment refer to the frequency of the LN chip mounted on the experimental holder. In the free-standing situation, the chip plate presented a higher frequency as represented in the plot by an arrow and the legend "pre-stress". e) Snapshot of the cross-section of FEM simulated displacement map for the mode $M_1$ of the chip partially covered by a layer of ice, as in the experiments. f) Comparison between the experimental $\Delta f/f_0$ vs MSD and a selected region of a FEM simulated $\Delta f/f_{free}$ vs $MSD_{free}$ curves (see supporting information section S11).*

**2.6 Critical assessment of the implementation of TSM-AWs for ice detection.**

Results in previous sections have revealed a distinct $S_{11}$ spectral evolution for icing experiments conducted under static or dynamic conditions. In the former case, we followed the freezing of sessile droplets; in the latter, the progressive layer-like accretion of ice on the surface of a LN plate. The two icing processes have been monitored following the evolution of a $S_{11}$ resonance peak previously identified as a TSM-AW resonant mode with considerable shear character (i.e., mode $M_1$). Focusing on this mode, the FEM simulation of the AW interaction with either water or ice aggregates on the surface (c.f. Figure 3) has shown that mechanical coupling of this acoustic mode is highly efficient for ice, but is unable to excite the bulk of a water droplet. A direct consequence of this property in the context of the present work is that the employed LN chips, albeit their simple layout (i.e., excited through two cm-scale extended electrodes at the edges of the plates), will not be very sensitive to the accumulation of water onto their surface, but will effectively react to the accumulation or the formation of ice.

In the static experiments conducted with one or more water droplets, the freezing led to the emergence of a complex, variable, and most often poorly reproducible peak pattern (see the color plots in Figures 4 and 5). For practical applications, this behavior confirms that the LN chip can detect phase changes in water droplets but makes it challenging to establish a clear correlation between the observed post-freezing $S_{11}$ spectra and experimental features such as droplet volume, the distribution of multiple water droplets on the surface, and other related variables. We deem that the complexity of the physical processes involved in the experiment precludes a clear and predictable evolution of the peak pattern profiles. Among these processes, we can quote the release of the latent heat of fusion during the freezing process (i.e., the heat associated to the phase change [51,56], the mechanism of freezing either through a bulk mechanism starting from the water-LN interface or a homogenous contour solidification mechanism affecting the droplet perimeter [42], the possible formation of small crystallites inside the water, the release of air bubbles or the increase in volume occurring after the water-ice transition, among others. All these physical processes will be much affected by variables of the experiment, such as temperature, cooling rate, humidity, size, and wetting angle of droplets, temperature gradients between the surface of the chips, and the surrounding air, etc. In particular, although the generation of surface strain has been scarcely addressed in the literature dedicated to droplet freezing, evidence exists that solidification may induce a significant stress at the substrate-ice interface and,[42,57,58] therefore, a certain substrate strain in and around the contact zone with the ice droplet [57]. We foresee that substrate-strained areas will strongly affect the AW transmission through the substrate and radically impair the resonance conditions of the LN plates. Experimentally, we have found that upon freezing large volumes of water or multiple smaller droplets, the resonance of the LN plate is almost suppressed, indicating that the resonant mode breaks down in the coupled system.

Despite these limitations, the experiment with just one droplet of water did provide relevant semiquantitative information regarding the evolution of the $M_1$ peak with temperature during the cooling stage (Figure 4 d) and the $M_1$* peak during the heating stage. We have attributed the same slope of the two linear trends to the effect of temperature on the electroacoustic properties of LN, as reflected in the temperature coefficient of frequency of this material [53]. From the perspective of sensing applications, the linear correlation can serve as an internal means to directly follow changes in temperature, an important asset for the use of LN chips in real-world scenarios [31].

The generation of local strain effects during ice monitoring is expected to take place mostly when ice accumulates progressively and homogeneously in the form of mixed ice on the surface of the LN plates, a situation rather similar to that encountered by the use of QCM to monitor the growth of thin films [38]. Our results in the IWT experiment following the progressive accretion of ice from the surface freezing of tiny supercooled water droplets impinging on the chip surface have shown that this process is accompanied by a systematic decrease in both the intensity and frequency of the $M_1$ peak (Figure 7). Compared to the water-droplet freezing experiment, it is anticipated that the gradual buildup of an ice layer does not induce comparable strains on the LN plate and that, as a first approximation, the evolution of its resonant properties primarily reflects the accumulation of ice mass on its surface. In this regard, the curve in Figure 7 d) evidences a clear dependence between the amount of ice accreted mass per surface area and the resonance frequency of the plate and therefore sustains the possibility to follow the time evolution of the icing process as expected for a reliable ice sensor. Interestingly, the evolution of frequency with the MSD of accumulated ice follows a profile that can be fitted with an exponential function. This fitting is advantageous when plotting experimental results on a logarithmic scale, since the results can be fitted with a straight line defined by just two points (see supporting information Figure S10).

The FEM simulation results for the ice accretion experiment under dynamic conditions (c.f. Figures 7 e and f) show that the TSM-AWs mechanically interact with the homogenous layer of ice formed in this experiment and that this interaction induces a change in frequency depending on the MSD following a polynomial curve with an "accelerated zone" that fits well with the shape of the experimental curve.

From an application point of view, it is also interesting that the $\Delta f/f_0$-MSD experimental evolution in successive accretion experiments was reproducible after de-icing the transducers using AW activation in a similar way to that of the experiment with the water droplets in Figure 6. This reproducibility clearly opens the way for the development of outdoor ice sensors for real scenarios, intercalating sensing intervals with short de-icing periods using AWs.

A final remark concerns the use of LN plates for practical ice-sensing applications compared with similar devices. Self-supported MEMS resonators [37], also using BAW, would likely provide a smaller limit of ice detection (i.e., higher sensitivity), but they would hardly withstand the harsh conditions existing in very demanding fields such as aviation. We deem that the relatively simple manufacturing of the chips and their proven robustness and sensitivity to ice, are advantageous factors of the proposed technology for the development of icing sensor systems.

## 3. Conclusions

A key conclusion from the reported results and discussion is that TSM-AW devices are particularly effective for the selective detection of ice, while they are less effective for detecting or monitoring changes affecting liquid droplets. This feature can be advantageous for developing ice-sensitive monitoring systems, as interference from liquid water would be minimal. Furthermore, devices based on LN plates with extended electrodes offer clear advantages over more complex transducers that use interdigital electrodes (IDTs) to excite other acoustic wave modes. These advantages include their simplicity, the ability to create chips in a wide range of shapes and sizes, and the easy manufacturing of electrodes. From a functional perspective, the relatively high Q factor of the $S_{11}$ resonant peaks in LN plates provides high accuracy for the detection of water-to-ice transformations, as well as excellent sensitivity to ice accumulation under dynamic conditions.

The $S_{11}$ peak of a resonant mode, identified as having a shear character, along with its evolution under various icing processes and conditions, has been adopted as a key parameter for detecting and monitoring icing. Furthermore, since most experiments can be reliably analyzed using finite element modelling by COMSOL Multiphysics simulations, this further supports the credibility of these observations, providing a valuable tool for predicting the behavior of chip devices made from other piezoelectric materials and thicknesses, as well as featuring electrodes with different arrangements, sizes, shapes, or thicknesses. Thus, for any alternative chip architecture, the methodology and finite element definitions used for simulation in this work would be useful for the preliminary exploration of resonant modes in transducers with other topologies. This predictive capacity is due to that the TSM mode is solely dependent on the thickness of the piezoelectric plates and consequently most of the simulation parameters will be valid for other flat geometries.

We emphasize that having high versatility in defining chip layouts and morphology is crucial for adapting transducers to various applications, holders, and even environmental conditions. Additionally, it is vital to respond appropriately (i.e., by generating TSM-AW modes) to predefined operating frequency ranges, especially when these are dictated by the overall architecture (including electronic) of the complete sensor system.

Our results have also shown that monitoring the occurrence of freezing/thaw processes is straightforward following the $S_{11}$ signal with a VNA. However, quantifying the extent of water-to-ice transformation under static conditions, where one or several droplets are deposited on the device surface, is difficult to achieve with the proposed operating mode. We have tentatively attributed the differences in $S_{11}$ spectra across experiments to local surface strains on the piezoelectric plate arising from the volume increase upon the transformation of water droplets into ice.

A proper quantification of icing processes has been demonstrated under dynamic conditions, provided that ice accumulates progressively on the chip surface within a well-defined icing area. We deem that under these experimental conditions, the layer-by-layer growth of the ice agglomerate does not permit the accumulation of much stress at the system interface, and the LN plate behaves as a resonator with a progressive accumulation of ice mass. It can therefore be concluded that avoiding inhomogeneities in the accreted area and proceeding upon the progressive accumulation of ice, either directly or by the sudden surface freezing of very tiny airborne droplets of water impinging onto the surface (e.g., as it happens for glaze ice onto flying aircraft and in the IWTs), are requirements for a straightforward operation of the transducers. These two requirements are met when using a collimator to select a well-defined area for ice accretion on the chip surface. Using this experimental configuration permits a quantitative estimate of the amount of accreted ice and facilitates the development of families of ice sensors based on LN chips, in which both a piezoelectric plate and a collimator arrangement are key elements of the device.

A final point to stress regarding the proposed sensor configuration is the possibility of recovering the chip's initial bare state by de-icing using the same type of AW used for monitoring. This permits the smart, autonomous, and continuous monitoring of the icing process, just increasing the intensity of the AWs to induce de-icing. The procedure would be simple: increase the electroacoustic excitation intensity once the $S_{11}$ signal reaches given frequency and/or intensity values, pre-defined as sensor saturation thresholds.

## 4. Methods

#### 4.1 Chip manufacturing and icing monitoring procedure

LN single-crystalline plates (128º Y-cut, 0.5 mm thickness, 1.5x3 $cm^2$) were purchased from Roditi International Corporation Ld.). As a first option for the preparation of chips, Ti/Au extended electrodes (10/90 nm thickness, 0.7x1.5 $cm^2$)) were deposited by thermal evaporation on the two extremes of the face of the plates side down the surface where icing processes will take place. Since similar results were obtained when a bare LN plate was directly supported on a metal tape with similar dimensions, a direct plate-metal tape arrangement has been used for most experiments. In most cases, the weight of the plate was enough to ensure a proper electrical contact between the LN plate and the electrodes. In the case of the IWT experiments, where the chip (i.e. defined here as the plate plus electrodes) had to be placed vertically, a holder frame and four spring pins closed to the four corners of the plate were employed to gently fixed its position (see Methods 4.3.3 for further information). The electrical connection between the chip and a VNA (2 ports vector network analyser, SDR-Kits, VNA 3SE) was made with a coaxial cable connecting the chip electrodes and a single port of the VNA (note that although $S_{11}$ spectra are recorded here using a VNA, measurement of the magnitude $|S_{11}|$ by a scalar NA would be sufficient for practical ice monitoring applications). The plate zones where the electrical connection was established were the only mechanically restrained surface areas of the crystal, with the rest of it remaining free-standing. A schematic description of the experimental arrangement, including the electronic connections, transducer layout and holder fittings is presented in the supporting information, Figure S1.

To monitor the icing phenomena we followed, in the selected scanned region of frequencies, the evolution of a resonance mode previously identified as a TSM-AW. To optimize the scanning procedure, a first wide frequency sweep was performed in a range comprised between 3.1 and 3.75 MHz. Once resonance modes of interest were identified, the frequency sweep range was made shorter to ensure an acceptable acquisition rate of at least three $S_{11}$ scans per second. This rate allows all curves to have a frequency resolution of at least 0.1 kHz. The recording of the $S_{11}$ curves of bare plates also provided a measurement of the quality factor Q of the chip acting as a resonator. This factor plays a key role for the simulation of chip resonances and power losses. Details about the calculation of this parameter from experimental spectra and the way to incorporate losses effects in the simulation model are provided in the **supporting information S2**.

The power applied by the VNA to the chip device during icing monitoring was in the μW range, clearly insufficient to produce any melting or transformation of the ice formed or accumulated on the surface. A macro was built in the SDR-kits software to automate data acquisition, enabling the recording of $S_{11}$ curves in a continuous way. The set of curves recorded in each experiment was processed and analyzed with custom scripts based on MATLAB. Data consisted of a large set of $S_{11}$ spectra, which served as the basis for the analysis as described in the Results and Discussion section.

Previous to any icing experiment, a precise characterization of the resonance modes found in each chip was carried out to ensure that in all cases the same vibrational mode was selected for monitoring. This prelaminar characterization of the plate chips was deemed indispensable because the frequency and relative intensity of $S_{11}$ peaks can be affected by minute changes in thickness or homogeneity of the plates or some other effects related to the misalignment of electrodes with the plate crystallographic directions. The analysis of these effects has been described in section 2.2.

### 4.2. Simulation of vibrational modes of transducer chips

FE simulation using the COMSOL Multiphysics package version 6.3 was carried out to characterize the resonance modes appearing in the frequency region selected for analysis (i.e. considered ranges were in the interval comprised between ca. 3.2 to 3.7 MHz, depending on the chip and experiment). A detailed description of the procedure is reported in the **Supporting Information S3**. A 2D model of the transducer has been built with the same dimensions and electrode distribution as for the chips of the experiments. The use of a 2D model instead of a 3D model was adopted to enhance computational performance, under the assumption that there is a translation symmetry in the chip geometry along the y direction, i.e., parallel to the short side of the LN plates. This implies that the model neglects all gradients of potential in this direction, as well as all border effects coming from the long sides of the chip plate. This means that the model solves the equations only in the x-z plane. To simulate the behavior of the crystal, the system has been modelled as a linear elastic solid subjected to electrostatic activation, both properties coupled by the piezoelectric tensor of the material. For the simulation, we chose the mechanical and electrostatic modules of COMSOL, incorporating the piezoelectric interface. The electromechanical properties of the LN plate, which was rotated to reflect the crystal cut used in the experiments, as well as its anisotropy and linear-elastic and piezoelectric properties, were defined by the same mechanical and electrical parameter tensors used in ref [59], using the stiffness, piezoelectric, and permittivity tensor elements proposed in ref. [60]. The boundary conditions taken for both the solid mechanics and electrostatic modules are described in the **supporting information S3**. Possible mechanical damping effects coming from the deposited metal electrodes or the gentle contact between the plate and the copper tape were neglected. The FEs were designed with a rectilinear grid with first-order shaped elements, consisting of 5820 nodes and 61758 degrees of freedom. Computations at given frequencies with various mesh densities proved that resolution did not affect the results (not shown). The mechanical boundary conditions reproduced the experimental holding arrangement of the device (c.f., Figure S1), where chip faces are not mechanically restricted, with the only exception of the area close to the edges where in the experiment the plate contacts the external copper bands of the setup.

The effect of water/ice placed on the chip surface was simulated for either water/ice droplets or for a homogeneous ice layer, as formed during the ice accretion experiments in the IWT. Domains geometry and mesh for these two cases are shown in Figure S3. Simulations are intended to describe the coupling between the AW moving through the substrate and the ice/water placed on its surface. Magnitude of displacements (for ice layers accreted on the surface of the LN plate) and accumulated energy density per unit volume (for water or ice droplets) have been determined.

For the simulations, ice has been taken as an isotropic homogeneous elastic solid material, whose mechanical behavior has been modelled as for a linear elastic material. The parameters used were its density and elastic constants, with values taken from Petrenko and Whitworth [61] in the case of ice droplets. The density of ice was adjusted to a value of 0.8 g cm$^{-3}$ for the simulation of the IWT experiment where a kind of mixed ice is formed [54]. (note however that in TSM devices the parameter affecting the frequency is the accumulated mass per unit of area (MSD in our case), [55]. For both $LiNbO_3$ and ice, losses were modelled to avoid unrealistic resonances. Elastic and dielectric losses were included for the $LiNbO_3$, while only elastic losses were modelled for the ice domain (see **Supporting Information S2 and S3** for further information).

Water was simulated assuming its fluid character with specific viscosity and adiabatic bulk modulus. Water properties were extracted directly from the COMSOL library [62] and its physics simulated using the linear form of the Navier–Stokes equations in a small parameter expansion (termed "Thermoviscous Acoustics" in COMSOL Multiphysics). Density mechanical energies (shown directly as density energy in Figure 3) have been defined for all domains as the kinetic energy peak value of the harmonic solution. This definition avoids using the potential elastic energy since this one would be different from ice and $LiNbO_3$ due to the piezoelectric nature of the latter.

### 4.3. Icing Experiments

Static and dynamic icing experiments were carried out in two cooling chambers and in an IWT facility, respectively. All experimental set up were provided with environmental control devices to monitor humidity and temperature and special fixtures to adjust the holders and, if necessary, to control the chip temperature with a Peltier. Well-adjusted feedthroughs permitted the passage of connecting cables to plug in the data acquisition systems placed in the exterior. The icing processes were directly recorded using cameras for image recording. Various icing experiments have been carried out, each one following a specific experimental protocol as detailed next.

#### 4.3.1 Single droplet water/ice phase change in static conditions

A small icing chamber was used for water/ice phase transformation experiments with one droplet of water placed on the surface of the chip. It was a home-made modification of a temperature chamber of an OCA20 goniometer (DataPhysics) provided with cable feedthroughs and a cool nitrogen flow to ensure a humidity-free environment. A Peltier permitted varying the chip temperature from ambient temperature up to -20ºC, as checked with a retractile thermocouple inserted in the chamber and placed at the Peltier surface.

Experiments proceeded in the following way: a 4 µl droplet of water was dispensed at room temperature on the surface of the LN plate. Then, temperature was progressively decreased by the combined cooling action of the $N_2$ flow and the Peltier. In most experiments, freezing took place close to -17ºC. Once the water droplet was frozen, the temperature was very slowly increased until melting of the formed ice aggregate took place. All over the icing experiments $S_{11}$ spectra were recorded continuously with the VNA during the cooling and the heating stages (elapsed time between recordings of 300 ms). High resolution lateral images of the water/ice droplets were taken simultaneously with a high-resolution photo camera for the freezing-thaw cycles, the frequency values were taken from one over one hundred $S_{11}$ recorded curves. The cooling/heating ramps were set with a ramp of 1 ºC/min, requiring approximately 30 min to reach the freezing temperature of -17 ºC, while the freezing/melting transitions amounted to ca. one hour each.

#### 4.3.2. Multiple droplets Water/ice phase changes in static conditions

A bigger cooling chamber was used to follow the transducer response when six water droplets of 4 µl or a single larger volume of water (i.e.,> 1.5 ml) were placed on the surface of the transducer chip. The droplets formed an array of 2 rows and 3 columns in the central part of the crystal. The crystal and droplets were placed inside the chamber at room temperature. Afterwards, the temperature was decreased with the chamber up to -10 ºC, making a smooth and controlled slow cooling ramp up to the working temperature. Along all this cooling process, the VNA was continuously monitoring the $S_{11}$ signal until the six droplets turned into ice. The main aim of this experimental approach was to determine the saturation condition of the transducer chip and its remnant sensitivity when ice is already present in some areas of the sensor.

#### 4.3.3 Ice accretion in dynamic conditions in IWT

These experiments were carried out in the icing wind tunnel described in ref. [45] under the following conditions: liquid water content (LWC) 0.5 $gm^{-3}$, wind velocity (V) of true air speed of 50 m $s^{-1}$ and median volume diameter (MVD) of 20 µm. To fix the position of the crystal, a special variation of the chip holder was employed, enabling to apply a gently pressure to the corners of the crystal by four small retractable springs (see Supporting Information 1). To properly select the accretion area of the ice on the surface of the chip and to ensure that accretion proceeds in a homogenous way, a rectangular collimator with a section of 0.3x1.3 $cm^2$ and a length of 1 cm was placed in front of the chip surface. [19]. From the values of LWC and V and the assumption that ice accretion rate is equivalent to the water droplet flux impinging onto the chip, it is possible to estimate the accreted amount of ice mass per surface unit, which is defined in this work as the Mass Surface Density (MSD). Therefore, after a given accretion time, t, this magnitude can be written as: $MSD(t)=LWC \cdot V \cdot t$. Assuming a density ρ, this relation permits to estimate an equivalent thickness of the ice layer, T, as $T(t)=MSD(t) \cdot \rho^{-1}$. "Ice thickness" is commonly taken as a reference parameter for many icing processes, particularly in aviation [54].

#### 2.3.4. AW de-icing of chips

Acoustic activation of the device was required to study its recovery possibilities in minus zero conditions. This was achieved by using a RF Generator (Siglent SSG3021 X) and a high-power amplifier (Mini-Circuits LZY-22+) to convey an RF power of 4 W to the device. Since the activation resonance band is considerably narrow, an oscilloscope (Siglent SDS1204X-E) was used to monitor the electrical activation of the chip, resetting the frequency manually in case of missing the resonant frequency. Additional details can be found in ref. [19].

## Acknowledgements

The authors thank the projects PID2022-143120OB-I00 and PCI2024-153451 funded by MCIN/AEI/10.13039/501100011033 and by "ERDF (FEDER)" A way of making Europe, Fondos NextgenerationEU and Plan de Recuperación, Transformación y Resiliencia. Project DGP_PIDI_2024_02239 funded by Consejería de Universidad, Investigación e Innovación of Junta de Andalucía and Project 202460E235 funded by CSIC. Project ANGSTROM was selected in the Joint Transnational Call 2023 of M-ERA.NET 3, an EU-funded network of about 49 funding organizations (Horizon 2020 grant agreement No 958174). The authors acknowledge the Wettability Characterization Service from ICMS for providing access to their facilities. The project leading to this article has received funding from the EU H2020 program under grant agreement 899352 (H2020-FETOPEN-2018-2019-2020-01 SoundOfIce).

***Supporting information S1. Scheme and picture of sample holder, transducer layout and electronic arrangements used for the icing monitoring experiments.***

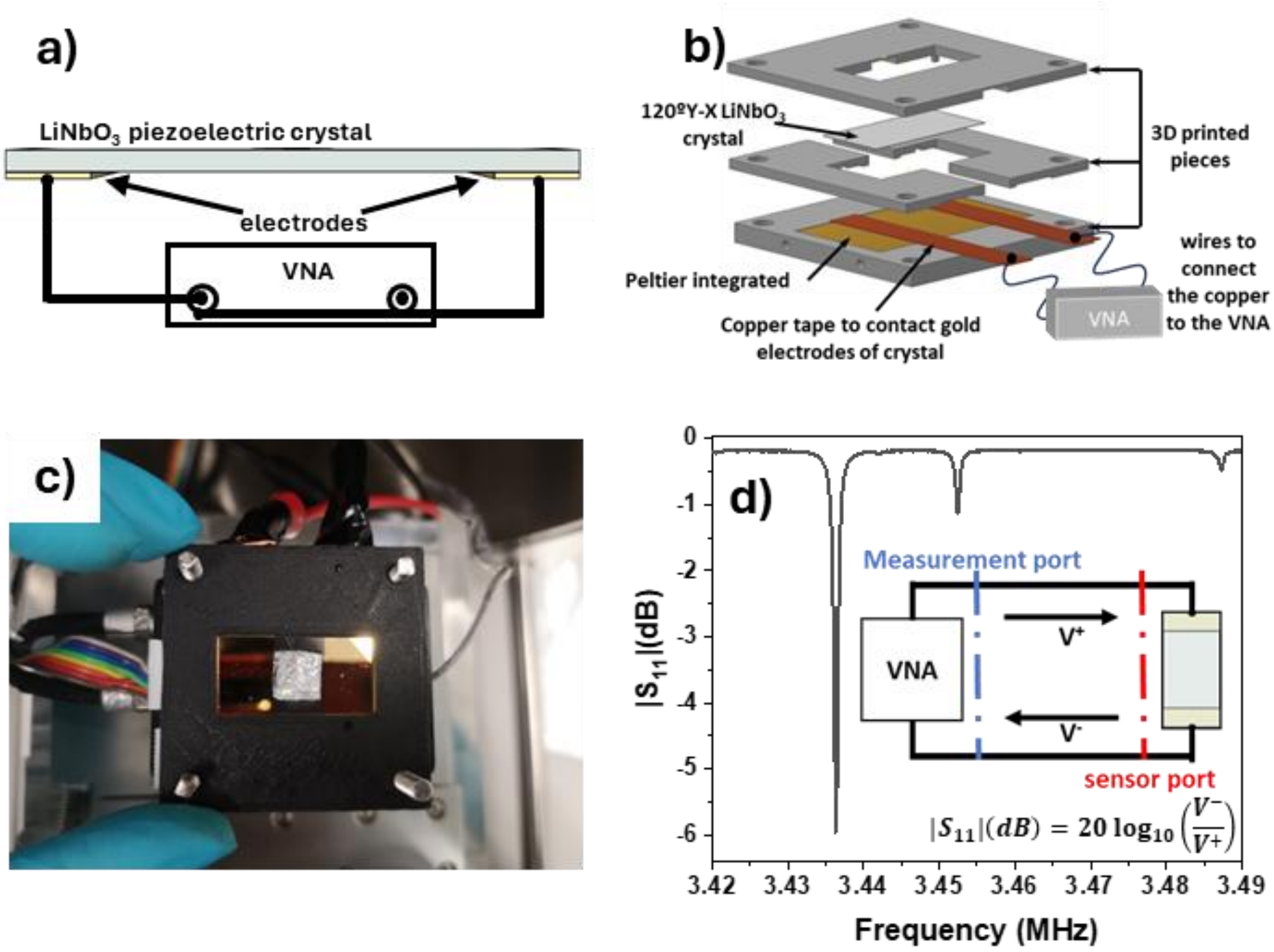

***Figure S1**. a) scheme of transducer chip layout in cross-section. b) Scheme of the sample holder showing the pieces integrated in the sensor case device. c) Picture of the chip device integrated in the holder case. d) Electronic arrangement used for the sensing measurements and $S_{11}$ signal monitored with the VNA in the frequency region selected for icing monitoring. The enclosed equation discloses the correlation of the $S_{11}$ signal and the voltages measured by the VNA*

The chip holder design in Figure S1 b) corresponds to the most common experimental configuration utilized in this work. On its base, two copper tape strips (0.4 mm thickness) are placed contacting the back surface area of the plates according to the layout in a). Outside the holder case, the copper tape stripes were contacted to a pair of wires using a crocodile-BNC coaxial cable, which was eventually connected to one of the ports of the VNA as indicated in d).

It should be noted that the holder was provided with a Peltier plate that was incorporated and used in the static experiments with one droplet of water (i.e., experiment whose results are summarized in Figure 4). Heat exchange between the Peltier and piezoelectric plate was rather efficient via the copper strips that contacted both.

***Supporting information S2. COMSOL simulations of viscous losses calculations using the Q factor derived from experimental data.***

COMSOL Multiphysics simulation of the different vibrational modes renders quite large unrealistic lattice displacements if no losses are included in the simulations. Although the relative displacement profiles were not affected by potential losses, the actual values decreased substantially when these are considered. A first approach to estimate these losses relies on the assessment of the Q factor of the resonant signals (i.e., $S_{11}$ resonance peaks). The Q factors have been determined from the experimental curve as $Q=f_0/$ FWHM, where $f_0$ is the main frequency of the $S_{11}$ peak and FWHM is its Full Width at Half Maximum. These two parameters are directly determined by fitting the experimental $S_{11}$ with a Lorentz curve (see Figure S2). The Q parameter can be taken as the ratio *Energy in/Energy dissipated.* Since we use harmonic solutions, this parameter would reflect the amount of energy dissipated per cycle. The energy losses can be related to viscous dissipation processes, which can be defined in the form of a viscous ratio as $\eta=1/Q$ [1]. These losses were incorporated in COMSOL by adding a complex part to the elastic tensor $C_{ij}$ as defined below in S3. The effect of viscosity has been taken as an isotropic contribution, i.e., just one single value is needed. The value estimated for the experimental $S_{11}$ peaks in Figure S2 is $\eta=1.773\cdot10^{-4}$.

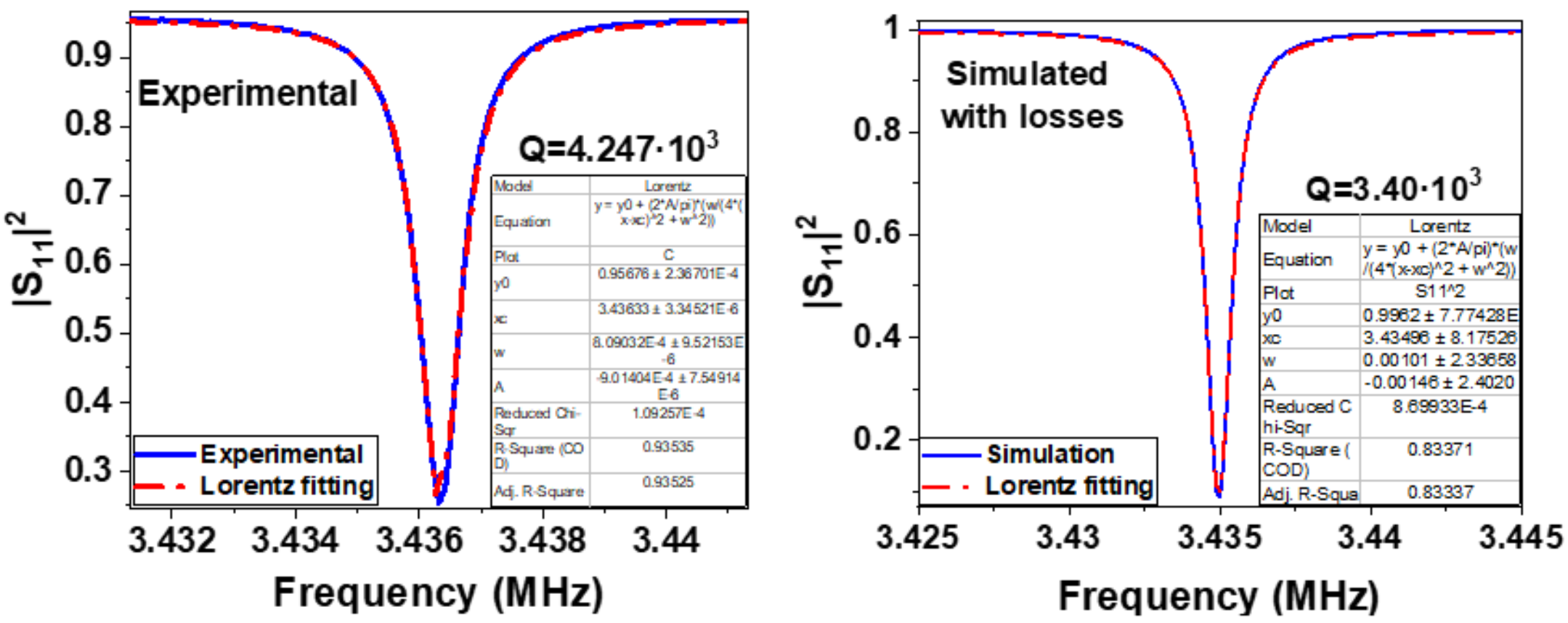

***Figure S2.*** *Plots of the experimental (left) and simulated(right) S11 signals considering the Q factor determined experimentally. Lorentz fitting curves are included in the analysis. The square insets summarize main parameters derived from the fitted curves.*

## *Supporting information S3. FEM implementation details*

In this section, we provide further details on the FEM simulations employed. Visualizations of the geometry and mesh implemented in the model are shown in Figure S3.

Simulation general specifications**:**

2D model working in the frequency domain, assuming time harmonic solutions.

·Geometry and mesh definitions:

The crystal is defined as a rectangle, as shown in Figure S3a where boundaries are also indicated (see below a detailed description of the zones between boundaries and their features regarding the simulation analysis). The plane selected for this 2D simulation is defined by an x-axis parallel to the long side of the crystal and a z-axis in the direction of the thickness of the crystal plate (i.e., perpendicular to the surface). Dimensions have been defined in accordance with the real crystal geometry. The mesh employed for the crystal domain, shown in Figure S2b, has been defined as a rectangular grid to better fit the shape of the crystal. On top of the crystal, dedicated domains have been defined for ice or water, whose geometry depends on the type of experiment being studied. Therefore, there are three different FEM geometries according to three different experimental conditions: iced droplets (or ice aggregates), liquid droplets and a rectangular-shaped ice layer. In the two former cases, water and ice have been modelled as hemispherical domains (Figure S3c), while the latter seeks to replicate the geometry of the accreted ice in the IWT experiment. In this last case, the same mesh element dimensions used for the plate were applied to the ice domains (Figure S2d). Meanwhile, hemispherical domains have been meshed using a free triangular mesh. Element size was adjusted to ensure that there are 22 elements per wavelength.

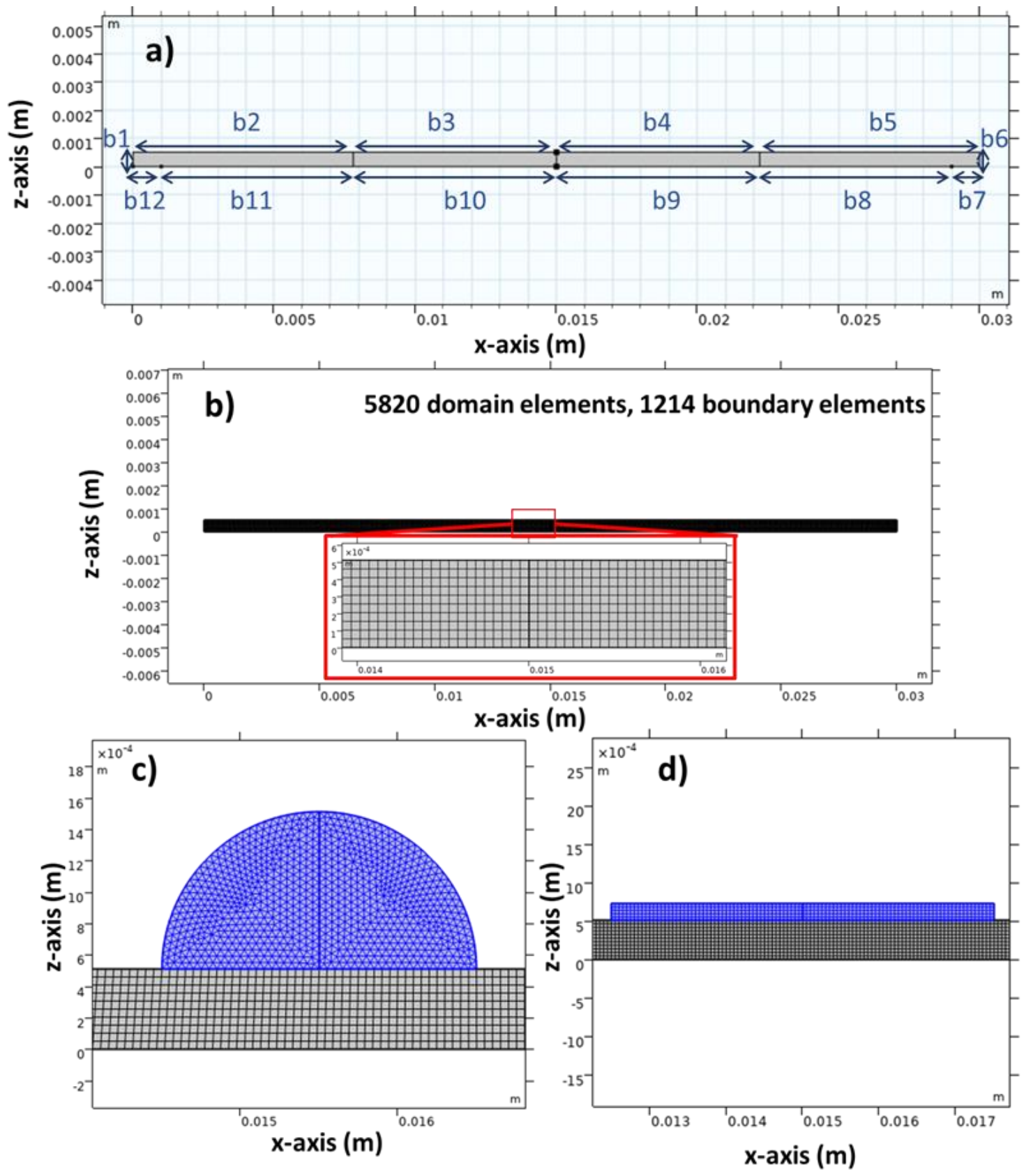

***Figure S3 Geometry and mesh details of the implemented FEM****. a) diagram of the 2D geometry implemented in the COMSOL model. The different boundary elements are identified as "bn" where n is the identification number of each boundary element, b) diagram of the rectangular mesh implemented in the model, with an inset providing a zoom in of the central zone. c) and d) Mesh implemented in the simulated water/ice droplets and the ice layer grown in the IWT experiments, respectively.*

Applied equations to all solid domains:

Equation (1) has been applied to all solid domains

$$D_i = \epsilon_{ij} E_j + e_{ij} S_{jk} \quad (1)$$

$$T_{jk} = c_{jklm} S_{lm} - e_{ijk} E_i \quad (2)$$

$$S_{ij} = \frac{1}{2}\left(\frac{\partial u_i}{\partial x_j} + \frac{\partial u_j}{\partial x_i}\right) \quad (3)$$

The magnitudes $D_i$, $E_j$, $S_{jk}$, $T_{jk}$ and $u_i$ are the tensorial components of, respectively, the electric displacement, electric field, strain, tension and solid displacement. Einstein summation convention of subindexes is assumed. The material properties used for the simulations are the dielectric permittivity ($\epsilon_{ij}$), piezoelectric coefficient ($e_{ij}$) and elastic constants ($c_{jklm}$) of the $LiNbO_3$. Dielectric and viscosity losses have been introduced by adding an imaginary part to dielectric and elastic constants in the following form:

$$c_{jklm}=(1+i\eta)c_{jklm}^{lossless} \ \epsilon_{ij}=(1\text{-}i\gamma_{ij})\epsilon_{ij}^{lossless} \qquad (4)$$

Where $\eta_s$ and $\gamma_\epsilon$ are the damping ratios due to intrinsic viscosity and dielectric losses, and i is the imaginary unit. Intrinsic viscosity losses are taken as the main cause of damping, while values from literature were used for the dielectric loss ratios a [2]. The intrinsic viscous ratio was defined as a single value for all elastic constants and was calculated from Q, the quality factor, extracted from the experimental results (see supporting information S2 for further details of the measured Q factors)

$$\eta=\frac{1}{Q} \qquad (5)$$

For the ice domains, only intrinsic viscous losses have been included, whose values have been taken from the literature [3].

· Applied equations to liquid domains (water):

The applied equations belong to the thermoviscous branch of the acoustic theory, where a perturbation approach has been employed, and the classical density and pressure linear relationship has been considered [4]. An adiabatic condition has been applied to the liquid domain contour. In their harmonic form, the corresponding equations are the following:

$$i\omega v_t=\nabla\sigma\sigma=\text{-}p_tI+\mu(\nabla v_t+(\nabla v_t)^T)\text{-}\left(\frac{2}{3}\mu\text{-}\mu_B\right)(\nabla\cdot v_t)I \qquad (6)$$

$$i\omega\rho_t+\nabla\cdot(\rho v_t)=0 \qquad (7)$$

$$\rho_t=\rho_0\left(\beta_Tp_T\text{-}\alpha_pT_t\right) \qquad (8)$$

Where, $\omega$,$\rho_t$, $v_t$, $\sigma$, $p_t$, $\mu$,$\mu_B$, $\rho_0$,$\beta_T$, $\alpha_p$ and $T_t$ are angular frequency, total density, total particle velocity, stress tensor, total pressure, dynamic viscosity, bulk viscosity, unperturbed (or base) density, isothermal compressibility, isobaric coefficient of thermal expansion, total acoustic temperature variation respectively. Material properties were taken directly from the COMSOL library.

·General applied boundary conditions:

The geometry of the crystal model is shown in Figure S3a where the different boundary elements are identified. The boundaries from b1 to b6, including b9 and b10 represents free tension faces of the crystal where no free charge is allowed to exist. The rest of the boundaries belong to the electrodes, and therefore, surface density free charges are there enabled, albeit they still behave as mechanical free traction boundaries. The only exceptions are b7 and b12 were outward displacements are not allowed because these boundaries take into account the crystal interaction with the experimental holder.

$$n_i D_i = 0\ ;\ \sigma_{ij} n_j = 0\ \ for\ \{[b1, b2, \dots, b6]U[b9, b10]\} \tag{9}$$

$$n_i D_i = \sigma_{free}\ ;\ \sigma_{ij} n_j = 0\ for\ \{b8, b11\} \tag{10}$$

$$n_i D_i = \sigma_{free}\ ;\ u_i n_i = 0\ for\ \{b7, b12\} \tag{11}$$

In equations (6-8) $n_i$, $D_i$, $\sigma_{ij}$, and $u_i$ are the components of the normal vector to each boundary, the electric displacement, the Cauchy stress tensor and the solid displacement vector, while $\sigma_{free}$ stands for the surface free charge on the surface, which is not directly defined, but simulated by the software as a harmonic excitation.

·Boundary conditions of ice boundaries

For the simulation of ice (In both droplet and rectangle shape, as shown in figures S3c-d), the outer boundaries are considered as mechanical free traction boundaries in the same way as previously defined for the crystal. Furthermore, the interface between ice and crystal has been considered as a perfect interface where no slip is allowed. This requires continuity of both the displacement field and the traction components of the stress tensors:

$$u_i^{crystal} = u_i^{ice};\ \sigma_{ji}^{crystal} n_i = \sigma_{ji}^{crystal} n_i \tag{12}$$

In expression (9), the subscripts indicate the domain of each magnitude.

· Boundary conditions of water boundaries

For the water domains, as seen in the previous section, a traction-free condition is applied to the outer boundaries of the droplet. However, at the interface between the water domain and the crystal, even if a no-slip condition is applied, the crystal and liquid velocity need to match. In the harmonic excitation form this requirement can be written as:

$$i\omega u_i^{crystal} = v_i^{fluid} \quad (13)$$

Electrical excitation definition:

The boundary elements belonging to the electrodes (b7, b8, b11 and b12) are modelled using the terminal node of COMSOL. The boundaries b11 and b12 behave as an electrical ground, while b7 and b8 are modelled as a terminated terminal in order to model the experimental behavior of the VNA. With the selected settings, the simulation model mimics a RF power source of 50 Ω of output impedance supplying 1 W of power to the electrodes port. Surface charge density is automatically controlled by the software.

***Supporting information S4.-Simulated $S_{11}$ curves for changes in the chip layouts (Euler angles) and variation of the thickness of the $LiNbO_3$ plates***

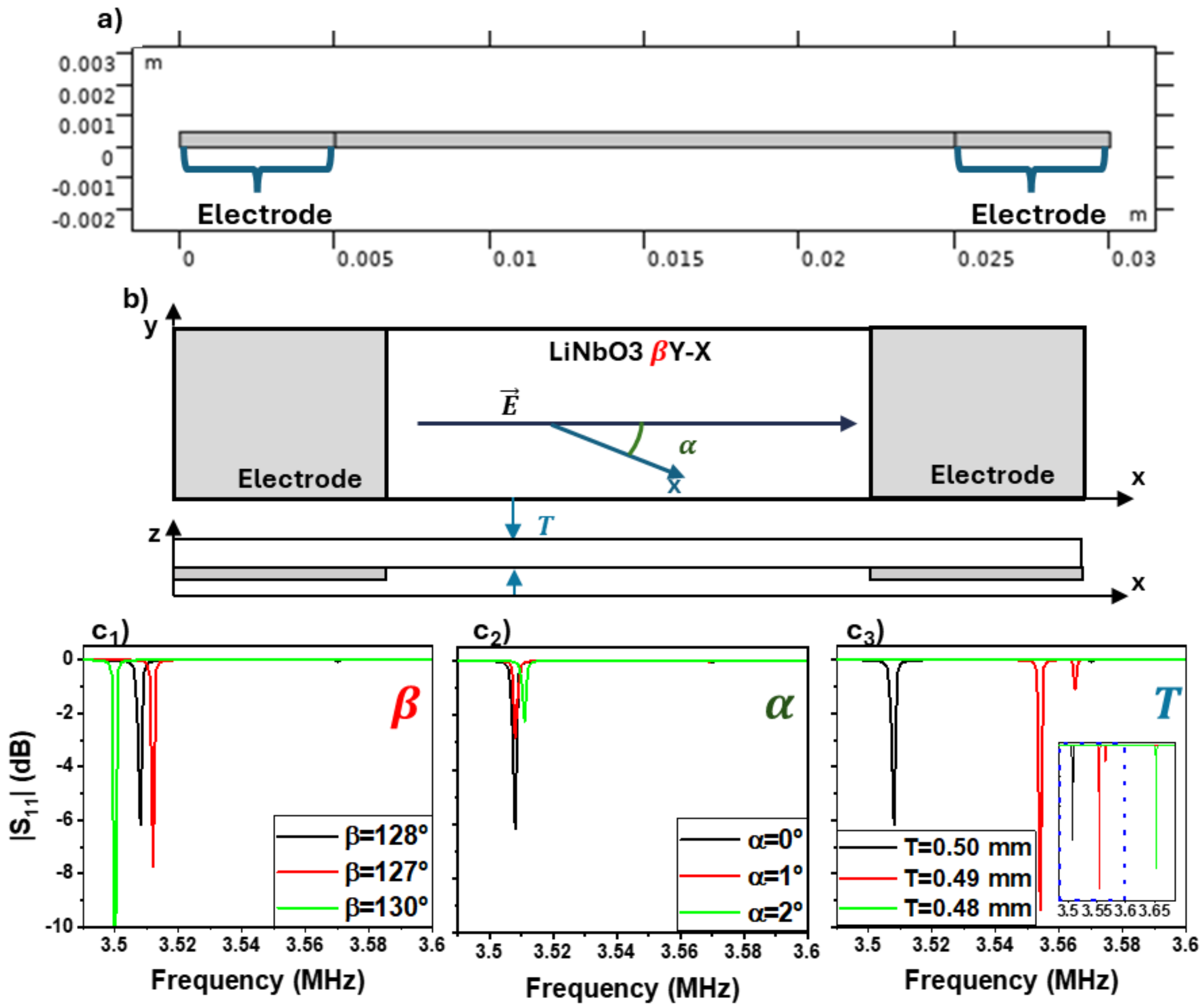

***Figure S4.*** *a) Scheme of the experimental arrangement of the chips (plate plus electrodes). b) Definition of Euler angles and plate thickness used as parameters for the calculations.* β *refers to any possible deviation*

*with respect to the nominal 128º cut of the crystal. c) Calculated spectra as a function of the values of angles* β *and α (c1, c2) and plate thickness (c3). The inset in c3 plotted over a large frequency range proves the strong influence of T on the resonant modes, showing that the $S_{11}$ main peak of the green curve (i.e. for a thickness of 0.48 mm) appears outside the frequency scale range selected for the panels.*

These simulations indicate that the $S_{11}$ resonance peak position is quite sensitive to the angle β and the plate thickness, but rather unsensitive to changes in α (i.e., to the misalignment of electrodes). Since the uncertainty of β is limited only to 0.15 º by the supplier, it is the thickness of the plate which has the highest impact on the resonant modes frequencies (certified nominal values within ± 30 um). Its influence is further explored in the main text and next in supporting information S5.

***Supporting information S5. COMSOL simulation of lattice vibrations modes along branches B1 and B2 for plate thicknesses around the transition region.***

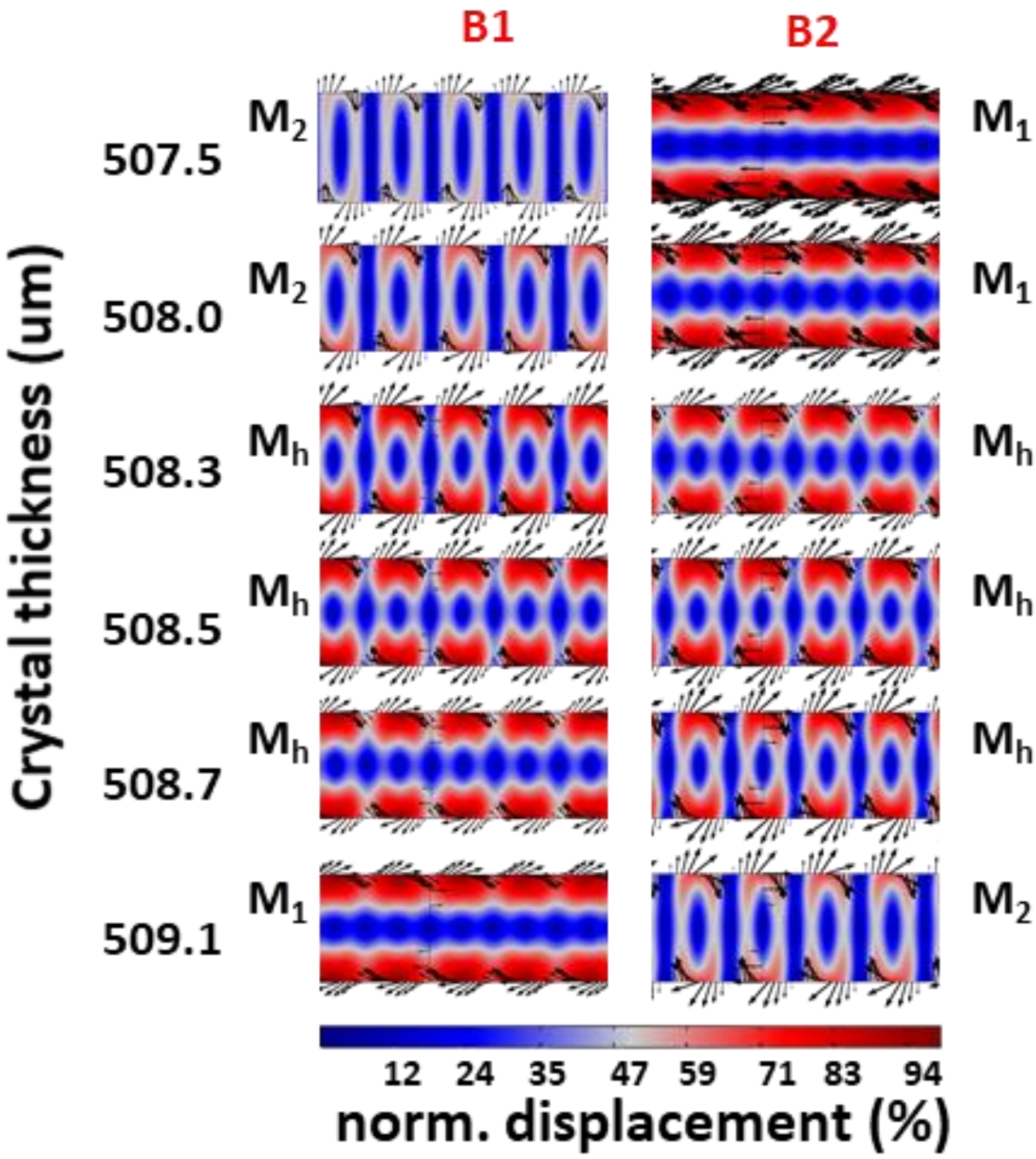

***Figure S5.*** *Snapshots of calculated COMSOL vibrational field as a function of thickness along branches B1 and B2 around the plate transition zone. Color scale is normalized and vectors at the surface represent the surface displacements in each case.*

The set of simulated snapshots reveals a change in the vibrational mode along branches $B_1$ and $B_2$ before, at, and after the transition zone. Branch B1 evolves from an anti-symmetric bending mode (i.e., mode $M_2$) for the thinnest plate to end up in a shear mode (i.e. mode M1) for the thickest examined plate. The reverse happens for branch $B_2$. For intermediate thicknesses, the vibrational modes along the two branches are quite similar, being both a hybrid mode between modes $M_1$ and $M_2$, named in this work as $M_h$.

## Supporting information S6. Discussion about the wavelength of wave components of $M_1$ mode

At a given time of the wave period (i.e. a certain phase shift of the harmonic solution) the surface polarization is dominated by vertical displacements giving rise to nodes separated by a $\lambda_x$, (general plane shift) with a majority bending component (the vertical ellipse in itself at the bottom graph of Figure 2c). This bending component is attributed to the contribution of an $A_0$ Lamb mode. This wavelength is different from that of the displacement in the z-axis inside the plate, $\lambda_z$ (not presented), which is dominated by the TSM component and has a value of two times the plate thickness, around 1020 µm. We hypothesize that the difference between $\lambda_z$ and $\lambda_x$ might be due to different phase velocities between the TSM waves and the $A_0$ Lamb waves. This is not extrapolated to $M_2$ and $M_3$, since displacement in the free-electrode region is only dominated by $\lambda_x$.

### *Supporting information S7. Chip characterization of the six droplet experiment*

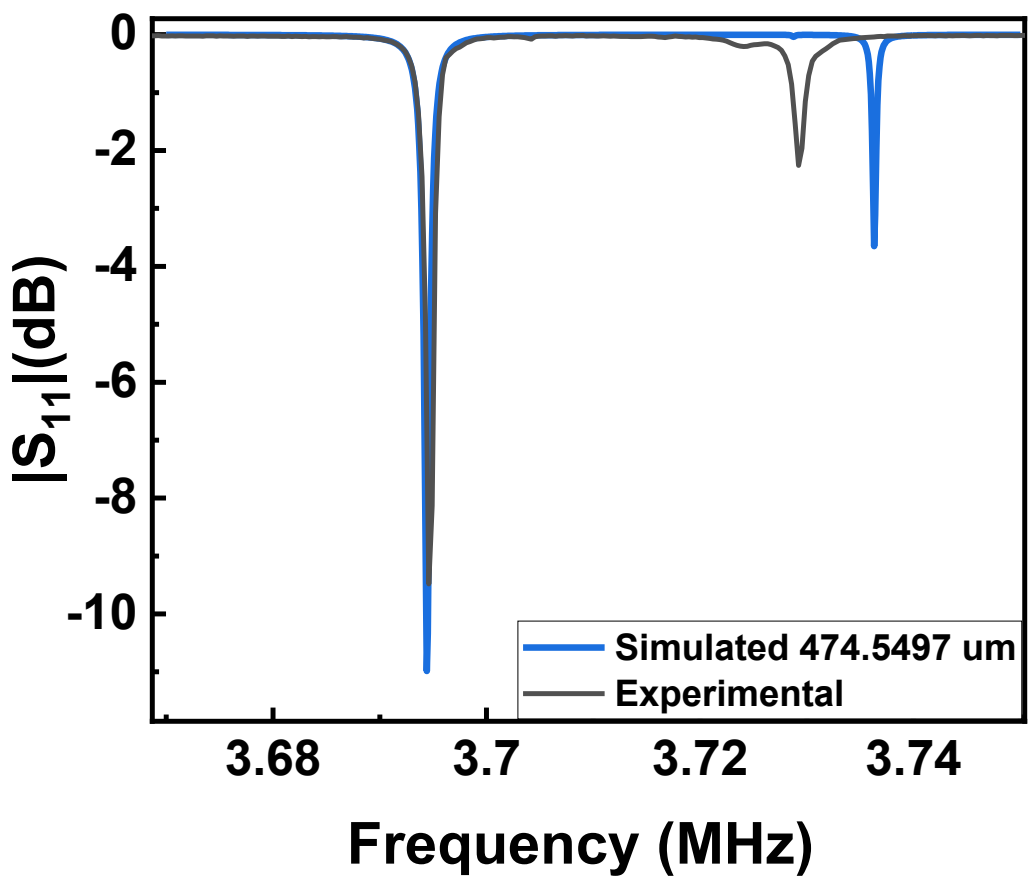

***Figure S7.*** *$|S_{11}|$ spectra of the pristine chip used for the six droplet freezing experiment and the COMSOL simulation of a chip plate with a thickness of 474.54 µm*

The chip used in the six droplet experiment shows modal frequencies that are considerably high (0.2 MHz above most chips used in this work). As it is discussed in the main text, chip resonances are mainly dominated by the thickness of the crystal plate. Thus, these higher frequencies are due to a thinner plate, as the simulations show in Figure S7. $M_1$ is well matched by the simulation, while a small discrepancy with the experimental data occurs for $M_3$, which stems from small differences in electrode extension between the real chip and the model.

***Supporting information S8. freezing detection of a big droplet deposited on the surface of the chip***

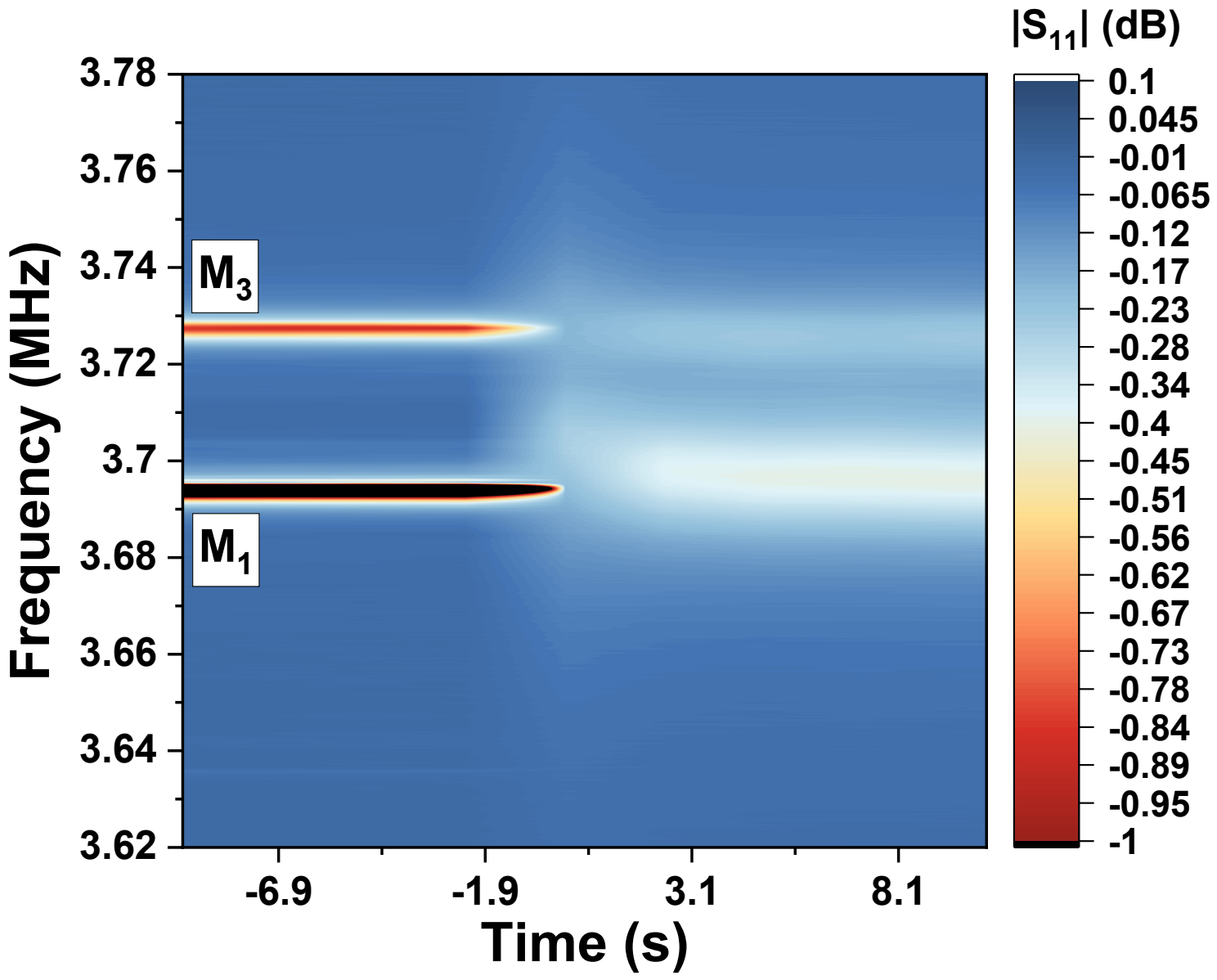

***Figure S8**. Colormap for a freezing experiment with a large droplet of 1.5 ml of volume. Time zero is defined as an arbitrary moment within the freezing transition. Only $M_1$ and $M_3$ modes appeared in this experiment, which has been attributed to the smaller thickness of the crystal used than that of the transition zone in Figure 3 of the main text.*

The colormap of Figure S8 clearly demonstrates the blurring of the observed resonance modes when a large volume of water is placed on the surface of the chip and transforms into ice. of an ice accretion experiment carried out in the IWT working with a LWC of 0.4 g $m^{-3}$. The inset shows the collimator dimensions and projected location onto the LN plate.

***Supporting information S9. Lateral image of the ice layer formed on the surface of the chip after an ice accretion detection experiment in the IWT***

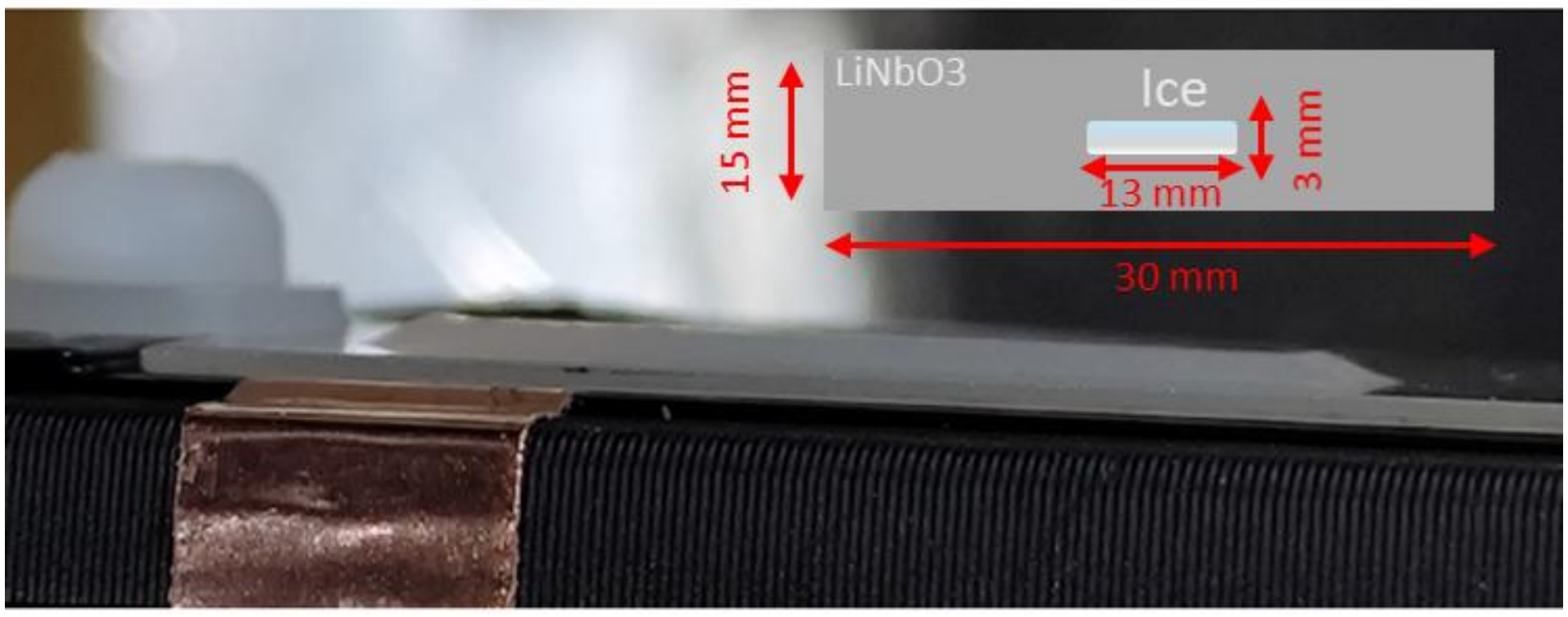

***Figure S9***. *Lateral enlarged image along the x-axis of the central part of the chip covered by a homogeneous and rectangular layer of ice. The image corresponds to the final stage of an IWT experiment.*

***Supporting information S10. Plot in a logarithmic scale of the $S_{11}$ relative frequency shifts in % as a function of the MSD for an ice accretion experiment carried out in the IWT***

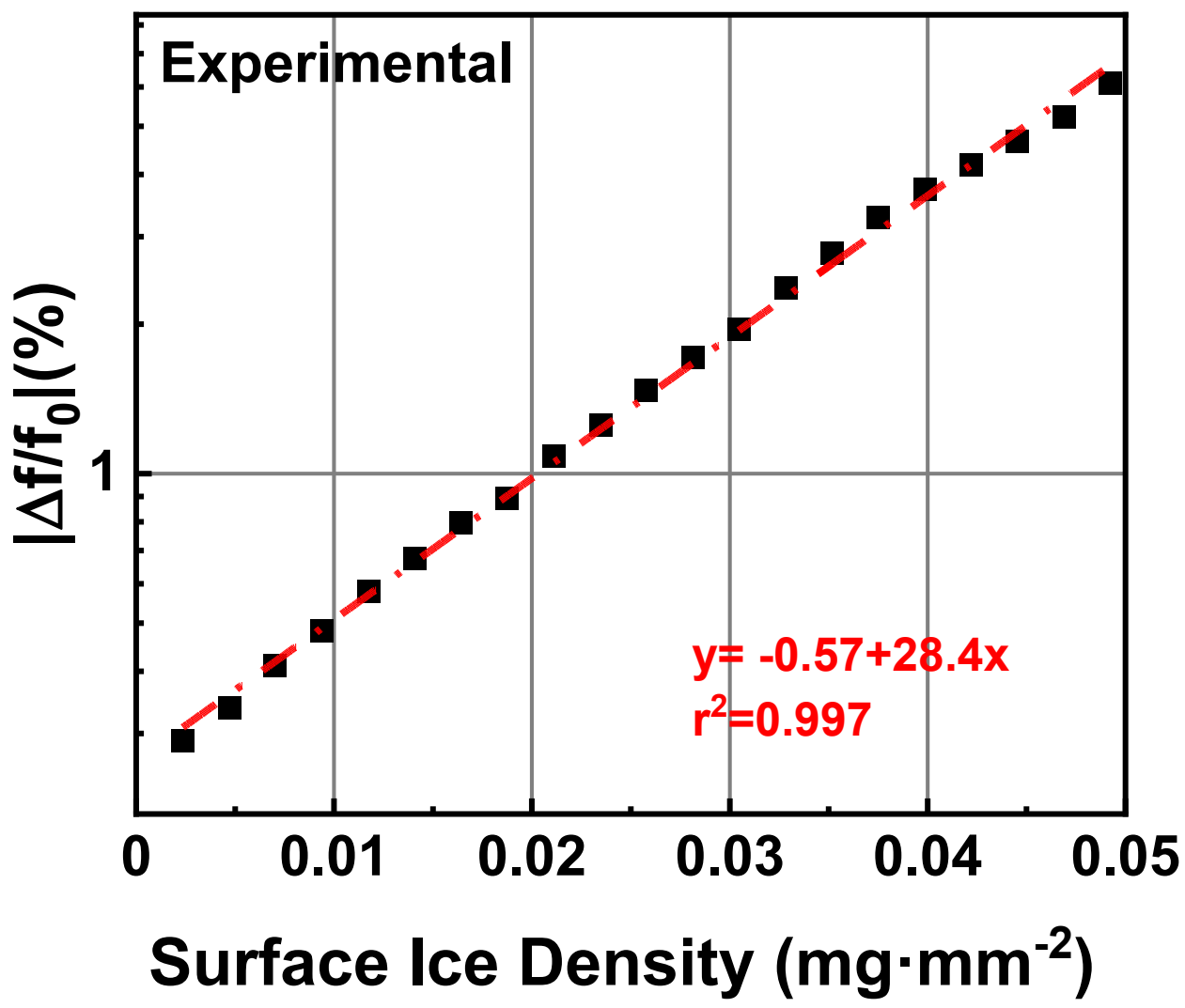

***Figure S10.*** *Logarithm representation of the relative frequency displacement factor in percentages versus the mass surface density (MSD) of accreted ice in an IWT experiment. Dots: experimental values. Red dashed line: fitting curve. The formula of the linear plot and the value of the regression coefficient ($r^2$) are included in the plot.*

***Supporting information S11. Simulation of the $S_{11}$ frequency shifts with the progressive accumulation of ice on a free-standing LN chip under the conditions of the IWT experiment.***

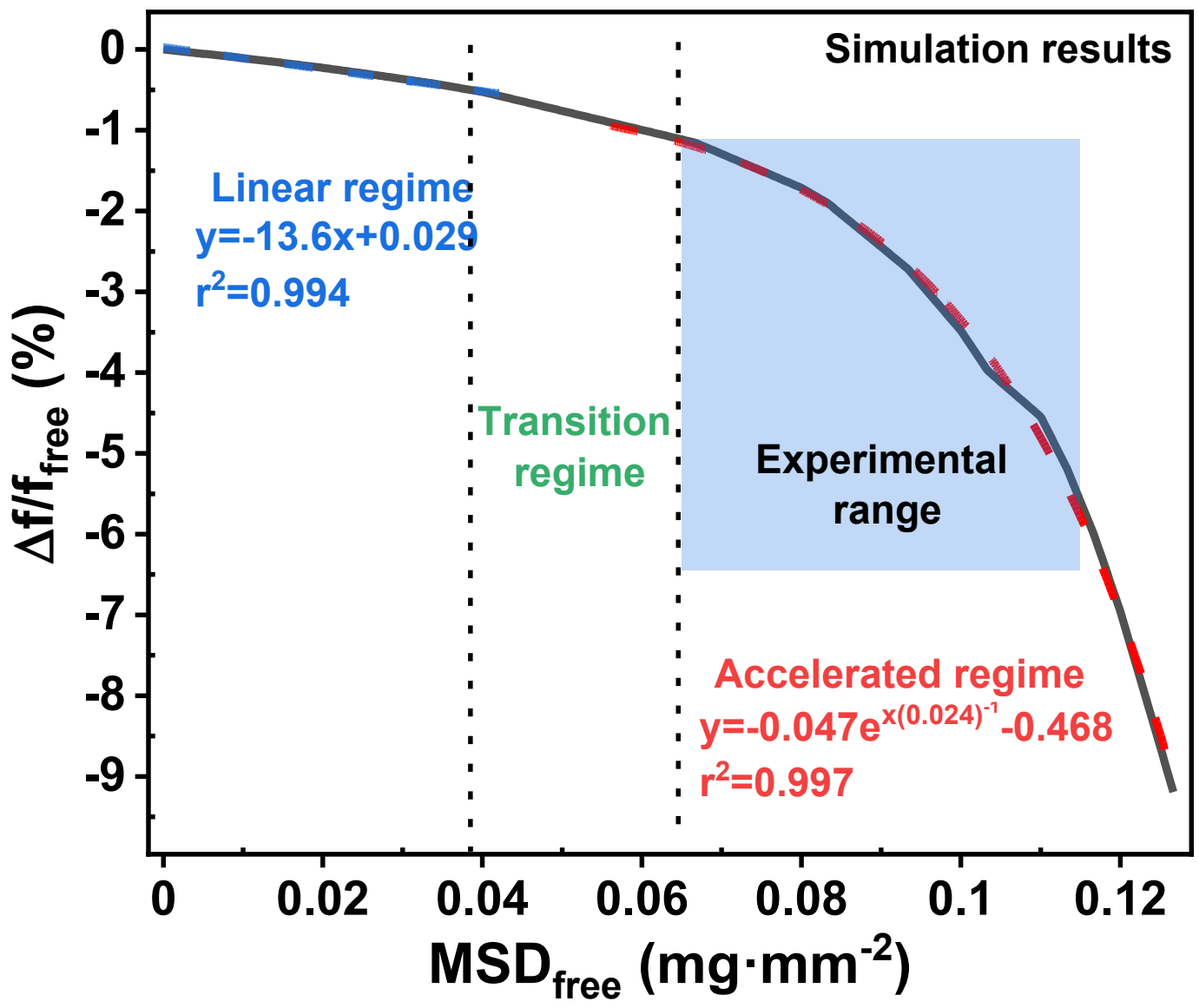

***Figure S11.*** *COMSOL calculation of the relative variation of vibrational frequency expressed in % for a free-standing LN chip with the $MSD_{free}$ of ice accumulated in the zone defined by the projection of the collimator in the IWT experiment*

Since the effect on the resonant frequency of the LN chips of the stress applied by the spring pins used to fix them in the IWT experimental set up is unknown, simulations were carried out for a free-standing plate and the same collimated accretion area than in the experiment (see method section of the paper). The obtained evolution of the $S_{11}$ frequency variation with the accumulated mass, expressed in terms of the $MSD_{free}$ is shown in Figure S10. The obtained curve has a polynomial shape whose initial part (i.e. up to approximately 0.04 mg mm$^{-2}$ can be fitted with a straight line (linear regime), followed by a sharply decaying curve (accelerated regime) that can be fitted with an exponential curve and adjusts well to the experimental decaying profile of the experiment (c.f., Figure 7f).

We hypothesize that the initial linear regime might be equivalent to the linear variation described by the Saubrey equation relating the Δf vs. Δm in the case of a QCM [5]:

$$\frac{\Delta f}{f_0} = -\frac{2 f_0}{A\sqrt{\mu_Q \rho_Q}} \Delta m \qquad (14)$$

where $f_0$ is the natural vibrational frequency of the QCM without any accumulation of external mass onto its surface, A is the active vibrating area and $\mu_Q$ and $\rho_Q$ the shear modulus and density of quartz.

***Video S1. Evolution of a droplet of water upon a freezing-thaw cycle***

***Video S2. Evolution of an ice aggregate during melting by AW activation***